\documentclass[doublespacing]{elsart}
\usepackage{icarus}

\usepackage{natbib}
\bibpunct{(}{)}{;}{a}{,}{,}
\usepackage{graphicx}

\begin{document}

\begin{frontmatter}
\begin{center}

\title{The Size Distributions of Asteroid Families in the SDSS Moving Object Catalog 4}

\author[label1,label2]{A. Parker}
\author[label1]{\v{Z}. Ivezi\'c}
\author[label3]{M. Juri\'c}
\author[label4]{R. Lupton}
\author[label5]{M.D. Sekora}
\author[label1]{A. Kowalski}

\address[label1]{Department of Astronomy, University of Washington, Seattle, WA 98195, USA}
\address[label2]{Department of Astronomy, University of Victoria, Victoria, BC V8W 3P6, Canada}
\address[label3]{Institute for Advanced Study, 1 Einstein Drive, Princeton, NJ 08540, USA }
\address[label4]{Princeton University Observatory, Princeton, NJ 08544, USA}
\address[label5]{Applied and Computational Mathematics, Princeton University, Princeton, NJ 08544, USA}

\end{center}
\end{frontmatter}

\date{}

\begin{abstract}
Asteroid families, traditionally defined as clusters of objects in orbital
parameter space, often have distinctive optical colors. We show that the separation 
of family members from background interlopers can be improved with the aid
of SDSS colors as a qualifier for family membership.  Based on an $\sim$88,000
object subset of the Sloan Digital Sky Survey Moving Object Catalog 4 with
available proper orbital elements, we define 37 statistically robust asteroid 
families with at least 100 members (12 families have over 1000 members) using 
a simple Gaussian distribution model in both orbital and color space. The
interloper rejection rate based on colors
is typically $\sim$10\% for a given orbital family definition, with four
families that can be reliably isolated only with the aid of colors. 
About 50\% of all objects in this data set belong to families, and this fraction
varies from about 35\% for objects brighter than an $H$ magnitude of 13 and rises to
60\% for objects fainter than this. 
The fraction of C-type objects in families decreases with increasing $H$ magnitude for $H > 13$, while the fraction of S-type objects above this limit remains effectively constant. This suggests that S-type objects require a shorter timescale for  equilibrating the background and family size distributions via collisional processing.
The size distribution varies 
significantly among  families, and is typically different from size distributions 
for background populations. The size distributions for 15 families display a 
well-defined change of slope and can be modeled as a ``broken'' double power-law. 
Such ``broken'' size distributions are twice as likely for S-type
familes than for C-type families (73\% vs. 36\%), and are dominated by 
dynamically old families. The remaining families with size distributions 
that can be modeled as a single power law are dominated by young families
($<$1 Gyr). When size distribution requires a double power-law model, the 
two slopes are correlated and are steeper for S-type families. No such 
slope--color correlation is discernible for families whose size distribution 
follows a single power law. For several very populous families, we find that 
the size distribution varies with the distance from the core in orbital-color 
space, such that small objects are more prevalent in the family outskirts. 
This ``size sorting'' is consistent with predictions based on the Yarkovsky
effect. 
\end{abstract}

\begin{keyword}
ASTEROIDS\sep ASTEROIDS, DYNAMICS\sep PHOTOMETRY
\end{keyword}

\newcommand\about {\hbox{$\sim$}}

\section{Introduction}

The size distribution of asteroids is one of most significant observational
constraints on their history and is considered to be the ``planetary holy 
grail'' (Jedicke \& Metcalfe 1998, and references therein). It is also one 
of the hardest quantities to determine observationally because of strong
selection effects. Recently, Ivezi\'{c} et al. (2001, hereafter I01) determined 
the asteroid size distribution to a sub-km limit using multi-color photometry 
obtained by the Sloan Digital Sky Survey (York et al. 2000; hereafter SDSS). 
Here we extend their work by using an updated ($4^{th}$) version of the SDSS 
Moving Object Catalog (Ivezi\'{c} et al. 2002a, hereafter I02a). 

The main goal of this paper is to study size distributions of asteroid
families. Asteroid dynamical families are groups of asteroids in orbital 
element space (Gradie, Chapman \& Williams 1979, Gradie, Chapman \& Tedesco 1989, 
Valsecchi {\em et al.} 1989). This clustering was first discovered by Hirayama 
(1918, for a review see Binzel 1994), who also proposed that families 
may be the remnants of parent bodies that broke into fragments. About half 
of all known asteroids are believed to belong to families;  
Zappal\'{a} {\em et al.} 1995 (hereafter Z95), applying a hierarchical
clustering method to a sample of 12,487 asteroids, finds over 30 families. 
Using the same method and a larger sample of $\sim$106,000 objects, 
Nesvorn\'{y} et al. (2005, hereafter N05) identify $\sim$50 statistically 
robust asteroid families. 

The size distributions of asteroid families encode information about 
their formation and evolution, and constrain the properties of the families' 
parent bodies (e.g., Marzari, Farinella \& Davis 1999; Tanga et al. 1999; Campo 
Bagatin \& Petit 2001; Michel et al. 2002; de Elia \& Brunini 2007; Durda 
et al. 2007; and references therein). Motivated by this rich information 
content, as well as the availability of new massive datasets, here we address 
the following questions
\begin{enumerate}
\item What is the fraction of objects associated with families?
\item Do objects that are not associated with families show any heliocentric
      color gradient?
\item Do objects that are not associated with families have uniform size
      distribution independent of heliocentric distance?
\item Do objects associated with families have a different size
      distribution than those that are not in families?
\item Do different families have similar size distributions?
\item Is the size distribution related to family color and age?
\end{enumerate} 

These questions have already been addressed numerous times (e.g.
Mikami \& Ishida 1990; Cellino, Zappal\'{a} \& Farinella 1991; 
Marzari, Davis \& Vanzani 1995; Z95; Morbidelli et al. 2003; N05). 
The main advantages of the size distribution analysis presented 
here, when compared to previous work, are 
\begin{itemize}
\item The large sample size: we use a set of $\sim$88,000 objects for which
      both SDSS colors and proper orbital elements computed by Milani \& 
      Kne\v{z}evi\'{c} (1994) are available
\item Simple and well-understood selection effects: the SDSS sample is 
      $>$90\% complete without a strong dependence on magnitude (Juri\'{c} et al. 
      2002, hereafter J02)
\item Improved faint limit: the sample of known objects listed in the latest ASTORB Þle from January 2008 to which SDSS observations are matched is now essentially complete to $r\sim19.5$ (corresponding to $H\sim17$ in the inner belt, and 
      to $H\sim15$ in the outer belt)
\item Improved family definitions due to color constraints (rejection
      of interlopers and separation of families overlapping in
      orbital space)
\item Improved accuracy of absolute magnitudes derived using SDSS photometry, 
      as described below.
\end{itemize}

The SDSS asteroid data are described in Section 2, and in Section 3
we describe a novel method for defining asteroid families using
both orbital parameters and colors. Analysis of the size distribution 
for families and background objects is presented in Section 4, and
we summarize our results in Section 5.

\section{  SDSS Observations of Moving Objects  } 

\subsection{     An Overview of  SDSS  }
\label{overview}

The SDSS is a digital photometric and spectroscopic survey which will
cover about one quarter of the Celestial Sphere in the North Galactic
cap, and produce a smaller area ($\sim$300 deg$^{2}$) but much deeper
survey in the Southern Galactic hemisphere (Stoughton et al. 2002; 
Abazajian et al. 2003, 2004, 2005; Adelman-McCarthy et al. 2006). 
SDSS is using a dedicated 2.5m telescope (Gunn et al. 2006) to provide 
homogeneous and deep ($r < 22.5$) photometry in five bandpasses 
(Fukugita et al.~1996; Gunn et al.~1998; Smith et al.~2002; Hogg et al.~2001; 
Tucker et al.~2006) repeatable to 0.02 mag (root-mean-square scatter,
hereafter rms, for sources not limited by photon statistics, Ivezi\'{c} et al.~2003) 
and with a zeropoint uncertainty of $\sim$0.02-0.03 mag (Ivezi\'{c} et
al.~2004). The flux densities of detected objects are measured almost
simultaneously in five bands ($u$, $g$, $r$, $i$, and $z$) 
with effective wavelengths of 3540 \AA, 4760 \AA, 6280 \AA, 7690 \AA, 
and 9250 \AA. 
The large survey sky coverage will result in photometric measurements for well 
over 100 million stars and a similar number of galaxies\footnote{The recent
Data Release 6 lists photometric data for 287 million unique objects observed
in 9583 deg$^2$ of sky; see http://www.sdss.org/dr6/.}.
The completeness of SDSS catalogs for point sources is $\sim$99.3\% 
at the bright end and drops to 95\% at magnitudes of 22.1, 22.4, 22.1, 
21.2, and 20.3 in $u$, $g$, $r$, $i$ and $z$, respectively. 
Astrometric positions are accurate to better than 
0.1 arcsec per coordinate (rms) for sources with $r<20.5$ (Pier et al.~2003), and 
the morphological information from the images allows reliable star-galaxy separation 
to $r \sim$ 21.5 (Lupton et al.~2002, Scranton et al. 2002). 
A compendium of other technical details about SDSS can be found 
on the SDSS web site (http://www.sdss.org), which also provides 
interface for the public data access.

\subsection{ SDSS Moving Object Catalog    }

The SDSS, although primarily designed for observations of extragalactic
objects, is significantly contributing to studies of the solar system objects 
because asteroids in the imaging survey must be explicitly detected and 
measured to avoid contamination of the samples of extragalactic objects
selected for spectroscopy. Preliminary analysis of SDSS commissioning data 
by I01 showed that SDSS will increase 
the number of asteroids with accurate five-color photometry by more than two orders 
of magnitude, and to a limit about five magnitudes fainter (seven magnitudes
when the completeness limits are compared) than previous multi-color surveys
(e.g. The Eight Color Asteroid Survey, Zellner, Tholen \& Tedesco 1985).
For example, a comparison of SDSS sample with the Small Main-Belt Asteroid
Spectroscopic Survey (Xu et al. 1995; Bus \& Binzel 2002ab) is discussed in
detail by N05.

\begin{figure}[H]
\centering
\includegraphics[width=12cm]{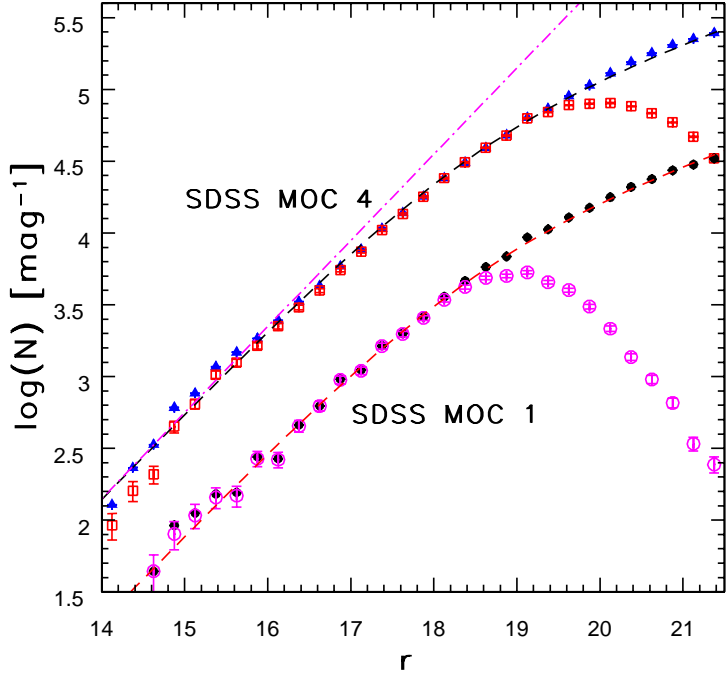}
% BB from 20 17 592 779 to 70 337 622 729
\caption{
An illustration of the improvements in sample size between the first and
the fourth release of SDSS Moving Object Catalog. Symbols with (statistical)
error bars show differential counts for moving objects listed in the first
release from 2002 (dots: all objects detected by SDSS; circles: identified
in ASTORB file) and the fourth release from 2008 (triangles: all SDSS;
squares: in ASTORB). Note that, in addition to a sample size increase
of about a factor of 7, the faint completeness limit for objects listed
in ASTORB also improved by a about a magnitude (the number of unique 
ASTORB objects increased from $\sim$11,000 to $\sim$100,000). The dashed 
lines show a double-power law fit described in text, with the $dlog(N)/dr$ 
slope changing from 0.60 at the bright end to 0.20 at the faint end. For
illustration, the dot-dashed lines shows a single power-law with a 
slope of 0.60.}
\label{counts}
\end{figure}

SDSS Moving Object Catalog\footnote{
% Available at
http://www.sdss.org/dr6/products/value\_added/index.html} (hereafter 
SDSS MOC) is a public, value-added catalog of SDSS asteroid observations
(I02a). It includes all unresolved objects brighter than
$r=21.5$ and with observed angular velocity in the 0.05--0.5 deg/day
interval. In addition to providing SDSS astrometric and photometric
measurements, all observations are matched to known objects listed 
in the ASTORB file (Bowell 2001), and to a database of proper orbital elements 
(Milani 1999; Milani \& Kne\v{z}evi\'{c} 1994)), as described in detail by 
J02. J02 determined that the catalog completeness (number of moving objects 
detected by the software that are
included in the catalog, divided by the total number of moving objects
recorded in the images) is about 95\%, and its contamination rate is about 6\% 
(the number of entries that are not moving objects, but rather instrumental artifacts). 
The most recent SDSS MOC 4th data release contains measurements for 471,000 moving 
objects. A subset of 220,000 observations were matched to 104,000 unique
objects listed in the ASTORB file (Bowell 2001). The large sample
size increase  between the first and fourth release of SDSS MOC is 
summarized in Figure~\ref{counts}. The object counts in both releases
are well described by the following function (I01)
\begin{equation}
 {\Delta N \over \Delta r} = n(r) = n_o {10^{ax} \over 10^{bx} + 10^{-bx}},
\end{equation}
where $x = r - r_C$, $a = (k_1+k_2)/2$, $b = (k_1-k_2)/2$, with $k_1$
and $k_2$ the asymptotic slopes of log($n$) vs. $r$ relations. This function
smoothly changes its slope around $r \about r_C$, and we find best-fit
values $r_C=18.5$, $k_1=0.6$ and $k_2=0.2$. The normalization constant, 
$n_o$, is 7.1 times larger for SDSS MOC 4 than for the first release.
In addition to this sample size increase, the faint completeness limit for 
objects listed in ASTORB also improved by a about a magnitude, to $r\sim19.5$ 
(the number of unique ASTORB objects increased from $\sim$11,000 to
$\sim$100,000). Above this completeness limit, the SDSS MOC lists color
information for $\sim$33\% of objects listed in ASTORB. 

The quality of SDSS MOC data was discussed in detail by I01 and J02, including 
a determination of the size and color distributions for main-belt asteroids. 
An analysis of the strong correlation between colors and the main-belt
asteroid dynamical families was presented by Ivezi\'{c} et al. (2002b, hereafter
I02b). Jedicke et al. (2004) reported a correlation between the family dynamical 
age and its mean color for S-type families, and proposed that it is due to 
space weathering effects. This correlation was further discussed and extended 
to C-type families by N05. Multiple SDSS observations of 
objects with known orbital parameters can be accurately linked, and thus 
SDSS MOC also contains rich information about asteroid color variability, 
discussed in detail by Szab\'o et al. (2004) and Szab\'o and Kiss (2008).

\subsection{ Errors in $H$ magnitudes listed in the ASTORB file    }

As pointed out by J02, there is a large systematic discrepancy between the 
absolute magnitudes listed in ASTORB file and values implied by SDSS
measurements. The latter are computed as 
\begin{equation}
\label{Hcorr}
                H_{corr} =  H_{ASTORB} + V - cV,   
\end{equation}
where $H_{ASTORB}$ is the ASTORB value, $cV$ is the apparent magnitude in
Johnson system computed from information listed in ASTORB as 
described in J02, and $V$ is the observed magnitude synthesized 
from SDSS $g$ and $r$ magnitudes (SDSS MOC entries 47, 42, and 32, 
respectively).  

\begin{figure}[H]
\centering
\includegraphics[width=12cm]{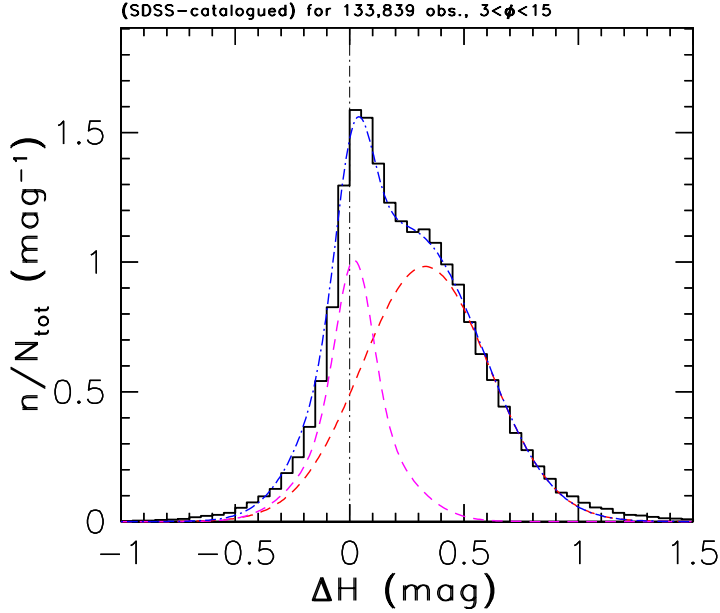}
% BB from 20 17 592 779 to 30 307 672 729
\caption{
A comparison of asteroid absolute magnitude, $H$, inferred from SDSS
measurements, and the value listed in ASTORB file for $\sim$133,000
observations of about 64,000 unique objects observed at phase angles
between 3 and 15 degrees. The histogram shows the data distribution
($\Delta H = H_{SDSS}-H_{ASTORB}=H_{corr}-H$, see eq.~\ref{Hcorr}) 
and the dot-dashed line is a best 
fit. The best fit is a linear combination of three gaussians: two (which simulate asteroid variability)
are centered on 0.02, have widths of 0.08 and 0.20 mag, and have 
relative normalizations of 13\% and 18\%, respectively. Their sum
is shown by the dashed line centered on $\Delta H=0.02$. The third
gaussian (which accounts for a large number of objects with bad photometry) has a width of 0.28 mag and is shown by the dashed line
centered on $\Delta H=0.33$. That is, about 69\% of $H$ measurements
listed in ASTORB file are systematically too bright by 0.33 mag.} 
\label{dH}
\end{figure}

This discrepancy persists in the $4^{th}$ release of SDSS MOC,
as illustrated in Figure~\ref{dH}. The mean difference between 
$H$ measured by SDSS and the values from ASTORB is 0.23 mag, and the
root-mean-scatter is 0.30 mag. The best-fit shown in Figure~\ref{dH} implies that uncertainty of $H_{corr}$ 
is about 0.16 mag, with a negligible systematic error (the latter
is expected to be about 0.02-0.03 mag due to uncertainties in 
absolute photometric calibration of SDSS imaging data; see
section 2.1). It is likely that this uncertainty
is dominated by magnitude variation due to rotation. The magnitude 
offset of 0.33 mag for $\sim$70\% of measurements implied by 
the best fit could be due to measurements reported by LINEAR.
A similar magnitude offset at the faint end is a known problem in
LINEAR calibration, and is currently being addressed with the aid 
of new calibration catalogs (J.S. Stuart, priv. comm.).

Since the random error in $H$ is twice as large as for $H_{corr}$, we 
adopt $H_{corr}$ in the remainder of this work. For a detailed
analysis of this magnitude offset problem\footnote{IAU Commission 15
has formed a Task Group on Asteroid Magnitudes to address this 
problem, see http://www.casleo.gov.ar/c15-wg/index-tgh.html},
we refer the reader to J02. 

\section{The Asteroid Families in SDSS MOC}

The contrast between dynamical asteroid families and the background 
population is especially strong in the space defined by proper orbital
elements. These elements are nearly invariants of motion and are thus well 
suited\footnote{The current asteroid motion is usually described by {\it osculating} orbital
elements which vary with time due to perturbations caused by planets, and are
thus less suitable for studying dynamical families.} for discovering objects 
with common dynamical history (Valsecchi {\em et al.} 1989, Milani \& 
Kne\v{z}evi\'{c} 1992).

\begin{figure}[H]
\centering
\includegraphics[width=8.5cm]{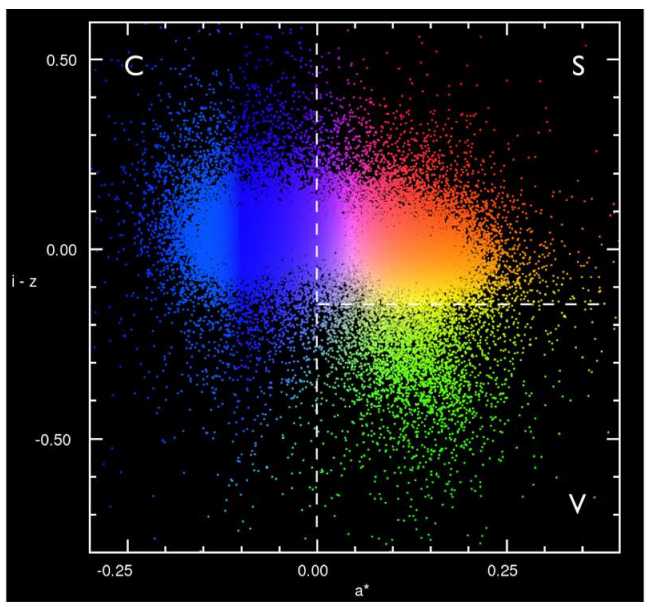}
\caption{
A plot of the color distribution in $a*$ and $i - z$ of 45,087 unique objects listed in 
both the SDSS MOC 4 and ASTORB file, and that have $H_{corr}<16$.
The approximate boundaries of three spectral classes are marked, and 
used in labeling family type. The color-coding scheme defined  here is 
used in figures \ref{prettyPlot1}--\ref{prettyPlot4}.}
\label{prettyPlot0}
\end{figure}

The value of SDSS photometric data becomes particularly evident when exploring 
the correlation between colors and orbital parameters for main-belt
asteroids. I02b demonstrated that asteroid dynamical families, defined as
clusters in orbital element space, also strongly segregate in color space. 
We use the technique developed by I02b to visualize this correlation for
$\sim$45,000 unique main-belt asteroids with $H_{corr}<16$ listed in SDSS MOC 4
(Figures~\ref{prettyPlot0}--\ref{prettyPlot1}). The asteroid color 
distribution in SDSS bands shown in Figure~\ref{prettyPlot0}, and its
comparison to traditional taxonomic classifications, is quantitatively
discussed by I01 and N05.

A striking feature of Figure~\ref{prettyPlot1} is the 
color homogeneity and distinctiveness displayed by asteroid families. In 
particular, the three major asteroid families (Eos, Koronis, and Themis), 
together with the Vesta family, correspond to taxonomic classes 
K, S, C, and V, respectively (following Burbine et al. 2001, we assume
that the Eos family is associated with the K class). Their distinctive 
optical colors indicate that the variations in surface chemical composition 
within a family are much smaller than the compositional differences between 
families, and vividly demonstrate that asteroids belonging to a particular 
family have a common origin.

\subsection{ A Method for Defining Families Using Orbits and Colors }

Traditionally, the asteroid families are defined as clusters of 
objects in orbital element space. The most popular methods for cluster 
definition are the hierarchical clustering and the wavelet analysis (Z95,
N05). Given the strong color segregation of families, it is plausible that
SDSS colors can be used to improve the orbital family definitions and minimize
the mixing of candidate family members and background population.

\begin{figure}[H]
\centering
\includegraphics[width=8.5cm]{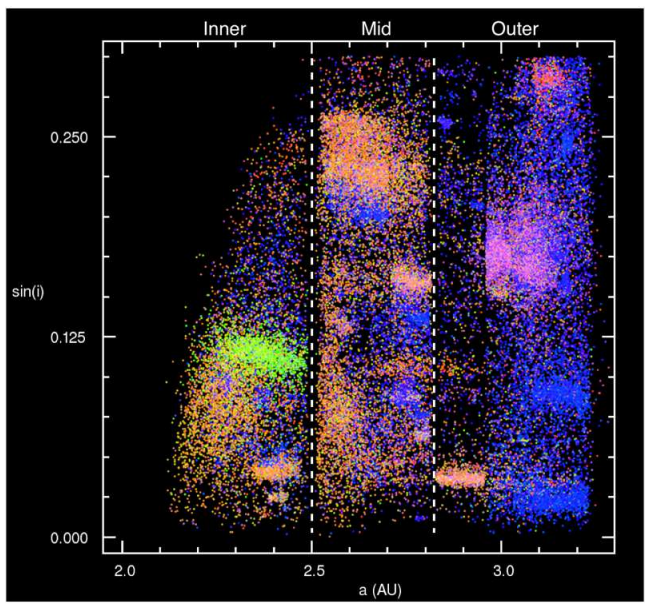}
\caption{%\scriptsize 
A plot of the proper $a$ vs. $sin(i)$ for the same objects as shown
in Figure~\ref{prettyPlot0}. The color of each dot is representative of 
the object's color measured by SDSS, according to the color scheme 
defined in Figure~\ref{prettyPlot0}. The three main regions of the belt, 
defined by strong Kirkwood gaps, are marked.}
\label{prettyPlot1}
\end{figure}

The SDSS colors used to construct Figures~\ref{prettyPlot0}--\ref{prettyPlot1}
are the $i-z$ color and the so-called $a^*$ color, defined in I01 as
\begin{equation}
              a^* \equiv 0.89(g - r) + 0.45(r - i) - 0.57
\end{equation}
The $a^*$ color is the first principal component of the asteroid color
distribution in the SDSS $r-i$ vs. $g-r$ color-color diagram (for 
transformations between the SDSS and Johnson system see Ivezi\'{c}
et al. 2007). Similar principal component analysis was also performed
Roig and Gil-Hutton (2006), who considered the distribution of taxonomic classes
(especially V-type asteroids) in SDSS principle components by comparing directly to 
spectroscopic data, and by N05, whose two principal components are well correlated with the 
$a^*$ and $i-z$ colors (we find that $a^\ast = 0.49\,PC_1 -0.16$ reproduces
the measured $a^\ast$ values with an rms of 0.026 mag for objects 
with $r<18$). The principal colors derived by N05
include the $u$ band, which becomes noisy at the faint end. Given
that the completeness of the known object catalog (ASTORB) reaches
a faint limit where this noise becomes important, we use the 
$a^*$ and $i-z$ colors to parametrize the asteroid color distribution. 
Therefore, the family search is performed in a five-dimensional
space defined by these two colors and the proper semi-major axis,
sine of the inclination angle and eccentricity. 

There are numerous techniques that could be used to search for clustering 
in a multi-dimensional space (e.g. Z95; N05; Carruba \& Michtchenko 2007). 
They differ in the level of supervision and 
assumptions about underlying data distribution. Critical assumptions 
are the distribution shape for each coordinate, their correlations, 
and the number of independent components. We utilize three different 
methods, one supervised and two fully automatic. The automatic unsupervised
methods are based on the publicly available code FASTMIX\footnote{See http://www.cs.cmu.edu/$\sim$psand}
by A. Moore and a custom-written code based on Bayesian non-parameteric 
techniques (Ferguson 1973; Antoniak 1974). 

In the supervised method (1) families are manually identified and modeled as orthogonal
(i.e. aligned with the coordinate axes) Gaussian distributions in orbital and color space.
The two unsupervised methods (2 and 3) also assume Gaussian distributions, but the orientation
of individual Gaussians is arbitrary, and the optimal number of families is determined by
the code itself. All three methods produce fairly similar results and here we describe only
the supervised method (1), and use its results in subsequent analysis. The two unsupervised
methods produce generally similar results for the objects associated with families, but tend
to overclassify the background into numerous (50-60) small families, and their details and
results are not presented thoroughly in this paper.

We select from the SDSS MOC 4 the first observation of all objects 
identified in ASTORB, and for which proper orbital elements are 
also available, resulting in 87,610 objects. Among these, 
there are 45,502 objects with $H_{corr}<16$. We split the main sample into 
three subsets using semi-major axis ranges defined by the major Kirkwood 
gaps (see Figure~\ref{prettyPlot1}): inner ($a<2.50$), 
middle ($2.50<a<2.82$) and outer ($a>2.82$) belt. For each subset, we 
produce the $e$ vs. $sin(i)$ diagrams color-coded analogously to
Figure~\ref{prettyPlot1}, and use them to obtain 
preliminary identification of asteroid families in both orbital and color 
space. Approximate rectangular bounds are assigned to these visually
identified families, from which median (centroid) and standard deviation,
$\sigma$, for the three orbital elements are estimated. Using these estimates, 
for each asteroid we compute distance in orbital space from a given family 
centroid as
\begin{equation}
\label{Dorb}
            D_{orbit} = \sqrt{d_a^2 + d_e^2 + d_i^2}
\end{equation}
where
\begin{equation}
      d_a = \frac{(a_{centroid} - a_{object})}{\sigma_a}
\end{equation}
\begin{equation}
      d_e = \frac{(e_{centroid} - e_{object})}{\sigma_e}
\end{equation}
\begin{equation}
      d_i = \frac{(i_{centroid} - i_{object})}{\sigma_i}
\end{equation}

Histograms in $D_{orbit}$ were used to determine a preliminary value of
$\Delta_{Orb}$, the maximum orbital distance from a family centroid 
for an object to be ascribed family membership. These initial $\Delta_{Orb}$
are determined from the differential $D_{orbit}$ distribution as the
position on the \emph{first local minimum} (the further rise of counts with
increasing $D_{orbit}$ is due to the background objects and other families).
The object distribution in the $D_{orbit}$ vs. $a^\ast$ and $D_{orbit}$ vs. ($i - z$)  
diagrams was used to first define approximate rectangular bounds for
each family, and then to compute the color centroid and standard deviation 
in $a^\ast$ and ($i - z$) for each candidate family. In order to use
color as a family discriminator, we define analogously to orbital elements
\begin{equation}
            D_{color} = \sqrt{d_1^2 + d_2^2}
\end{equation}
where
\begin{equation}
      d_1 = \frac{(a^\ast_{centroid} - a^\ast_{object})}{\sigma_a^\ast}
\end{equation}
\begin{equation}
      d_2 = \frac{(iz_{centroid} - iz_{object})}{\sigma_{iz}}.
\end{equation}
We use histograms in $D_{color}$ to define  $\Delta_{Col}$, the maximum 
color distance from a family centroid for an object to be ascribed family 
membership. Figure~\ref{dCol_dOrb} illustrates the $D_{orbit}$ and $D_{color}$ histograms
for the Vesta and Baptistina families, and the bottom panels show the distributions of family and background 
objects in the $D_{orbit}$ vs. $D_{color}$ plane.

In cases of families which formed from the disruption 
of a differentiated parent body, the color distribution
might not provide a well defined morphology. However, we did
not find any case where a subset of objects selected using 
$D_{orbit}$ did not result in one or two well defined color 
distributions. Nevertheless, it is possible that a small fraction
of objects could be rejected from a family due to different
colors than the majority of other members.

All objects that have both $D_{orbit} \leq \Delta_{Orb}$ and $D_{color} \leq
\Delta_{Col}$ are then considered to be a family member. With a given 
estimate of family populations, this procedure is iterated and all 
parameters are refined. It typically takes one to two iterations to 
converge. All ``converged'' families are removed from the sample and
the process was repeated until there were no family candidates with 
more than 100 members. This condition is the result of the requirement
that the statistical errors for the slope of absolute magnitude distribution of the families 
are smaller (typically 0.01-0.02) than plausible systematic errors
(0.03-0.04), as discussed below. 

Using this procedure, we found 37 families which account for 46\% of all
objects. Their defining parameters are
listed in Table 1. Additional three candidate families that had fewer than 
100 members in the last iteration were discarded (see last three entries in Table 2). The family names were 
determined by comparison with Z95 and
N05, and, when no corresponding family was found, by searching for the 
lowest-numbered asteroid in the Milani \& Kne\v{z}evi\'{c} (1994) catalog 
of proper orbital elements. In a small number of cases, it is possible 
that the ``name-giving'' object has a color that is inconsistent with 
the majority of objects in the family. We ignored such cases and retained
nomenclature from Z95 and N05 in order to ease the comparison. That is, the 
defining properties of families are the orbital and color parameters listed 
in Table 1, rather than their names.

\begin{figure*}[H]
\centering
\includegraphics[width=14cm]{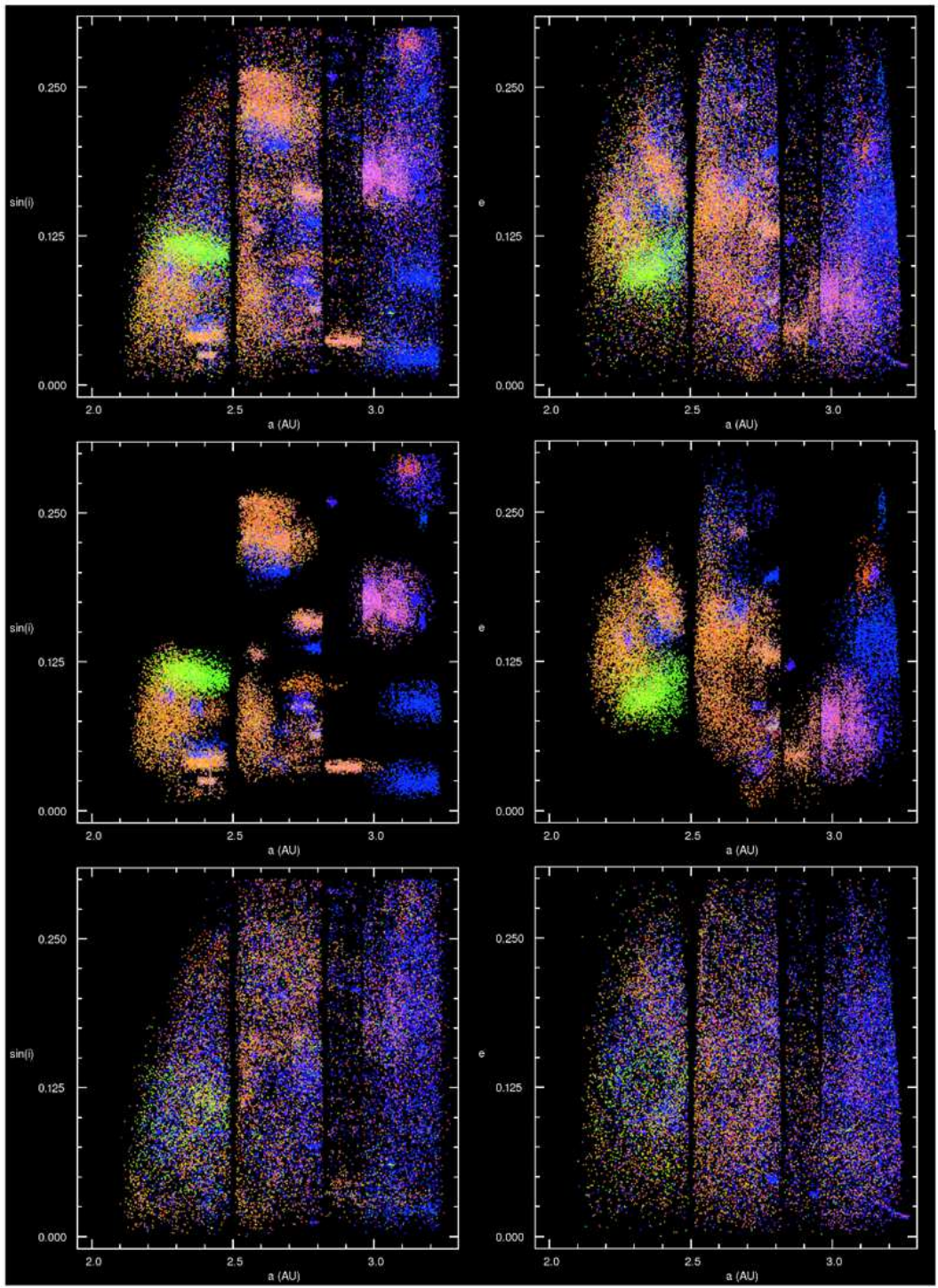}
\caption{%\scriptsize 
Illustration of the decomposition of the main-belt asteroid population
into families and background objects in proper $a$ vs. $sin(i)$ (left panels) and proper $a$ vs. $e$ (right panels). The top panels show all (background and family) objects in the data subset. The two middle 
panels show objects from 37 identified families, and the bottom two panels 
show the background population. }
\label{prettyPlot3}
\end{figure*}

\begin{figure*}[H]
\centering
\includegraphics[width=14cm]{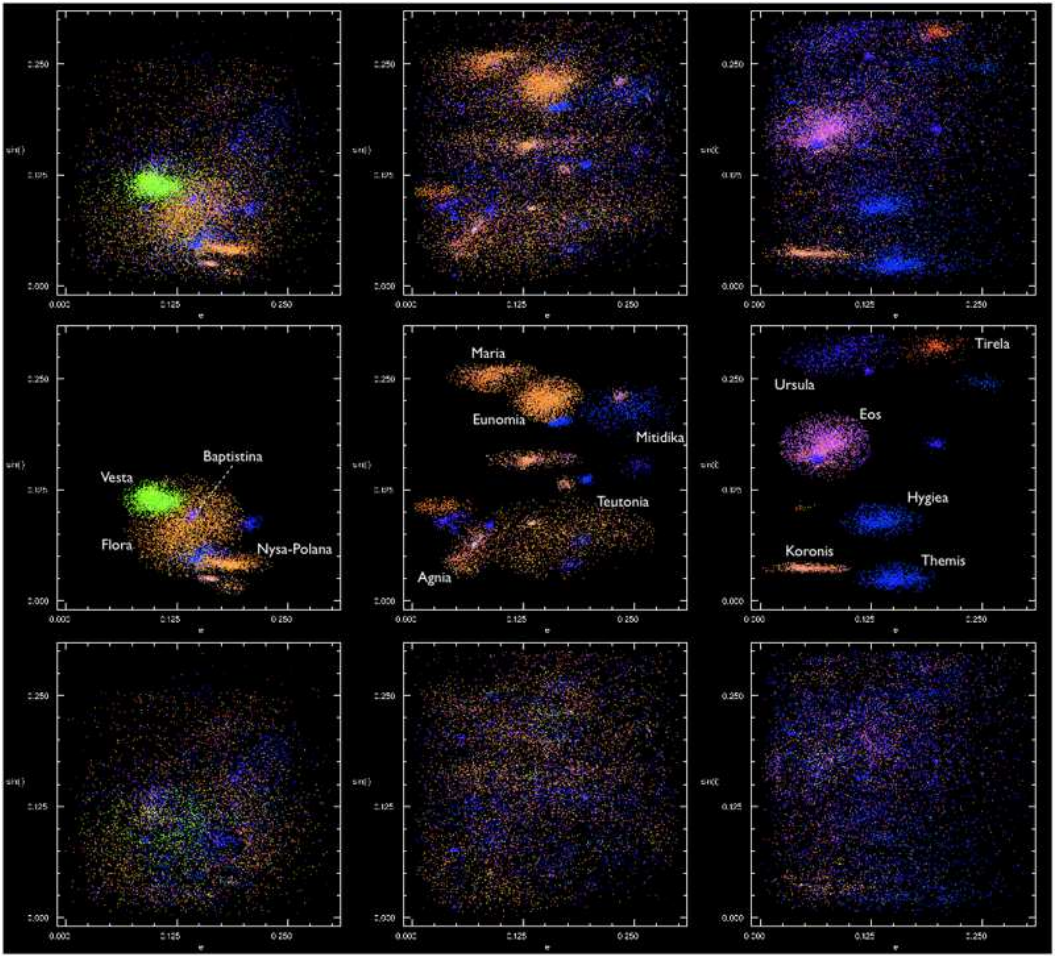}
\caption{%\scriptsize 
Analogous to Figures~\ref{prettyPlot3}, except that the top three panels 
show the $e$ vs $sin(i)$ distribution for the three main regions defined by
strong Kirkwood gaps ($a<2.5$ left, $2.5<a<2.82$ middle, $2.82<a<3.5$ right; 
see Figure \ref{prettyPlot1}). The middle row shows family members (with
several families of note labeled), and the bottom row shows the background population. For a high-resolution
version of this figure with complete labeling, see http://www.astro.washington.edu/ivezic/sdssmoc/sdssmoc.html}
\label{prettyPlot4}
\end{figure*}

The separation of the main-belt asteroids into families and the background
objects is illustrated in Figures~\ref{prettyPlot3} and \ref{prettyPlot4}. 
While there is some residual structure in the background, both in color and 
orbital space, it is much less prominent than the structure displayed by
identified families. 

\subsection{ The Comparison of Resulting Families with Previous Work }

Important questions about the quality of family associations derived here, 
that are relevant for subsequent analysis, are
\begin{itemize}
\item The contamination: are all families robust? 
\item The completeness: are any families missed?
\item What is the impact of variations in adopted family definitions on   
      the resulting family properties such as the number of members and 
      their size distribution? 
\end{itemize}
In this section we address the first two questions, and discuss the
third one below (\S \ref{syserr}). 

It is unlikely that the derived sample of 37 families contains any
spurious family because of the conservative requirement that a family
must include at least 100 members. Indeed, we have rejected three familes
that are likely real because thay had fewer candidate members (5, 90 and 46).
The very narrow color distribution of all selected families provides
another argument for their robustness. 

Our method of iterative removal of identified families and the simultaneous
use of colors and orbital parameters to search for remaining families 
appears very robust when compared to two automated methods also employed. 
Nevertheless, given that all three methods from
this work assume gaussian distributions of colors and orbital parameters,  
it is prudent to compare our family list with families obtained by other 
means, such as the hierarchical clustering method. We have cross-correlated 
the list of 37 families determined here with the list of 41 families obtained 
by N05 using the hierarchical clustering method. N05 based their study on a 
larger sample of objects with proper orbital elements ($\sim$106,000 vs. 
$\sim$88,000 analyzed here; note that the latter sample extends to $\sim$1.5 mag
fainter flux limit but is smaller because it includes only objects observed
by SDSS), and did not place a requirement on the minimum number of objects 
per family. Therefore, it is plausible that families missed by our selected method 
may be present in their list. 

Out of 41 families from the N05 list, 27 are listed in Table 2. This is 
encouraging level of agreement given the significant difference in 
applied methodology. We examined in detail each of the fourteen N05 families 
missing from our list and searched for them in the sample of background
objects. We did not find any candidate family that included more than 100 
members, though most appear to be real clusters. 

Among the ten families from our 
list that we could not identify in the N05 list, three were detected by at 
least one method discussed by Z95, and thus are likely real (Euterpe,
Teutonia, and Henan). Of the remaining seven, the recognition of four families 
was greatly aided by  color information (Baptistina from Flora, Mitidika and
Juno, Lydia and Padua, and McCuskey from Nysa-Polana). It is likely that the 
remaining three families were not detected by N05 because they have steep 
absolute magnitude distributions and thus only a small number of members 
were present in the (older) version of catalog used by N05. For example,
among the 3405 objects in the Teutonia family, only 37 have $H_{corr}<14$.

We further compared our list of families to those presented in Mothe-Diniz \emph{et al.} (2005), who used
spectroscopic measurements to probe asteroid family structure. Of their 21 nominal families, 
all but three (Renate, Hoffmeister, and Meliboea) can be matched to families detected here. 
Of the numerous smaller (many with fewer than 100 members) ``clumps'' they identify, 
10 correspond to families listed here, while eight families (Teutonia, Mitidika, Euterpe, Andree, 
Lydia, Ursula, Lyxaohua and Theobaldia) present in our list do not appear in Mothe-Diniz \emph{et al.} (2005).
 They resolved the remaining family in our data set, Flora, into a number of smaller clumps that merged
 into a single family at high cutoff velocities.

\begin{figure}[H]
\centering
\includegraphics[width=12cm]{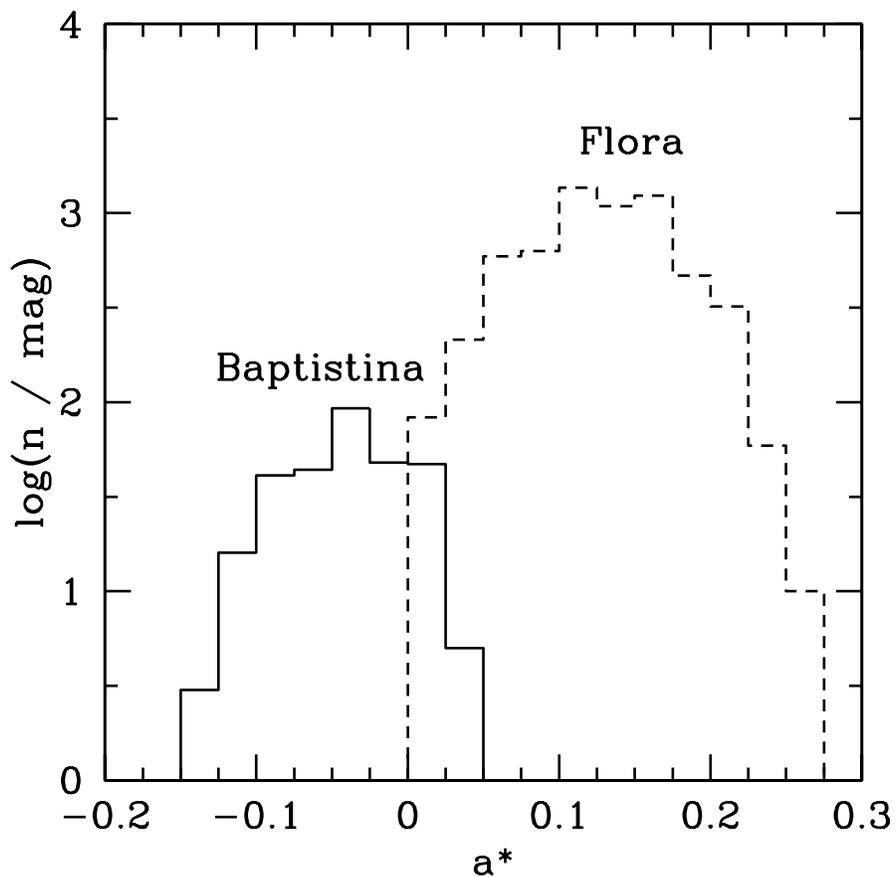}
% BB from 20 17 592 779 to 30 317 582 729
\caption{%\scriptsize 
An illustration of color differences for families with practically
identical orbital parameter distributions. The dashed histogram shows
the $a^\ast$ color distribution for 6,164 candidate members of the
Flora family. The solid histogram shows the $a^\ast$ color distribution
for 310 candidate members of the Baptistina family, which is easily 
separated from the Flora family thanks to the SDSS color information.} 
\label{colorHist}
\end{figure}

\begin{figure*}[H]
\centering
\includegraphics[width=6cm]{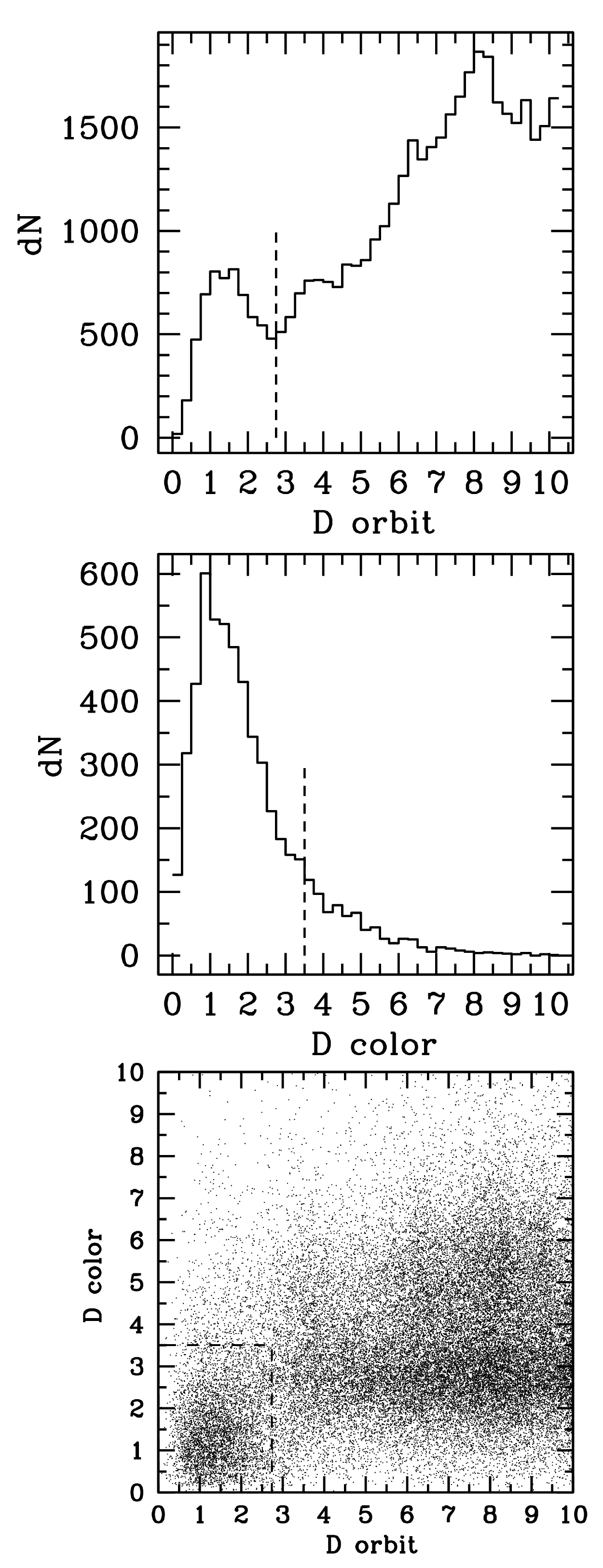}
\includegraphics[width=6cm]{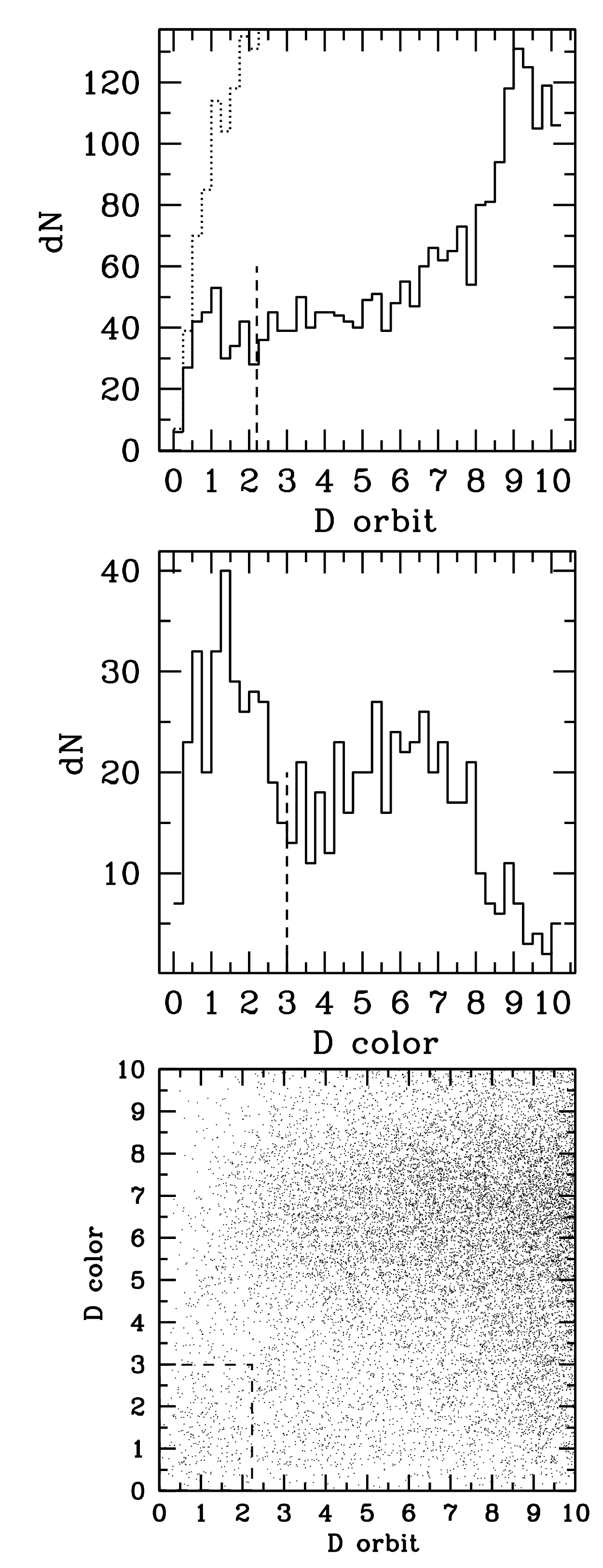}
%% BB 7 147 763 837
\caption{%\scriptsize 
\emph{Top Left:} $D_{orbit}$ histogram for the Vesta family and surrounding objects (defined from the Vesta family centroid). 
\emph{Top Right:} Equivalent to top left plot, but for the dynamically buried Baptistina family. The dotted line represents the $D_{orbit}$ histogram without any color constraints, which climbs smoothly to very high numbers of objects because of the inclusion of Flora family objects. The solid line represents the $D_{orbit}$ histogram with the color constraints applied, which displays a stronger clustering signature.
Vertical dashed line represents $\Delta_{Orb}$ cutoff values selected for these families.
\emph{Middle Left:} $D_{color}$ histogram for objects that met the $D_{orbit}$ criteria for the Vesta family.
\emph{Middle Right:} $D_{color}$ histogram for objects in a preliminary orbital definition of the Baptistina family, showing strong color distinction from the background Flora objects.
Vertical dashed line represents $\Delta_{Col}$ cutoff values selected for these families.
\emph{Bottom Left and Right:} $D_{orbit}$  vs. $D_{color}$ for Vesta and Baptistina families, respectively. Dashed box defines $\Delta_{Orb}$ and 
$\Delta_{Col}$ boundaries (objects inside are assigned family membership).
}
\label{dCol_dOrb}
\end{figure*}

A good example of the separation of dynamically mixed families using
SDSS colors is provided by the small family Baptistina which is ``buried''
within the Flora family. Figures~\ref{colorHist} and  \ref{dCol_dOrb} (right panels) illustrate how different 
$a^\ast$ color distributions enable the identification of $\sim$5\% of the objects nominally
assigned Flora family ($\langle a^\ast \rangle = 0.13$) membership by their orbital parameters as being members
of the Baptistina family ($\langle a^\ast \rangle = -0.04$). While initially puzzled why
a similar color-aided search by N05 did not yield any additional 
families, we have found that all color-separated families extracted here
are dominated by faint objects and thus may not have been present
in sufficiently large numbers in the older catalog. We conclude that
all the families discussed here are robustly detected, and that 
it is very unlikely that we missed any family with more than 100 
hundred members. It is, however, possible that the background is
composed of {\it numerous} families dominated by small objects 
that are not discernible with the presently available catalog. 
Hence, the fraction of $\sim50\%$ of objects associated with 
families is only a lower limit (this fraction is a function of object
size, as discussed below).

Figure~\ref{avgCol} shows the color dependence of the family and background populations
on semi-major axis. We note that the median $a^\ast$ color 
for the background population becomes bluer as the semi-major axis
increases in the same way as the median color for the family 
population.

\section{ The  Size Distribution of Asteroid Populations }

The known object catalog is complete to $r\sim19.5$; above this limit an ASTORB 
entry is found for practically every SDSS moving object. Depending on the
distance and orientation of the observed object, this apparent magnitude
limit corresponds to a completenes limit ranging from $H\sim17$ in the
inner belt to $H\sim15$ in the outer belt. Brighter than these limits,
selection function is essentially equal to 1 for the purposes of this 
work (SDSS managed to observe only about 1/3 of all ASTORB objects,
but this is essentially a random selection without an impact on derived
absolute magnitude distribution of individual families). These simple 
selection effects allow us to derive robust family size distributions
to very small size limits. The only other study of family size distributions
that approached the same size limit is that of Morbidelli et al. (2003),
who had to introduce an ad hoc indirect correction for selection effects 
in the known object catalog (which is supported by our analysis, as discussed
below). 

The transformation between the asteroid absolute magnitude, $H$, and its 
effective diameter, $D$, requires the knowledge of the absolute visual albedo 
$p_V$,
\begin{equation}
\label{absmag}
          H = 18.1 - 2.5 \log({p_V \over 0.1}) - 5 \log(D / 1 {\rm km}).
\end{equation}
The absolute albedo is not known for the overwhelming majority of objects 
in our sample. However, the albedo is known to be strongly correlated with
colors (Zellner 1979; Shoemaker {\em et al.} 1979; Delbo 2004); for example,
the C-like asteroids ($a^* < 0$) have a median albedo of 0.04 and the S-like 
asteroids have a median albedo of 0.14. Given that the color variations within 
a family are small,
it seems plausible that the albedo variation within a family is also small
(this is supported by the data compiled by Tedesco, Cellino \& Zappal\'{a} 
2005; see their Table 7). With this assumption, the {\it shapes} of the 
absolute magnitude distribution and the distribution of $\log(D)$ are the 
same. Hereafter, we will interchangebly
use ``the absolute magnitude distribution'' and ``size distribution'', where
the latter implies the distribution of $\log(D)$. For simplicity, in the
remainder of analysis we only use the differential distributions.

If the differential absolute magnitude distribution, $n(H)=\Delta N/\Delta H$,
can be described by
\begin{equation} 
                  \log(n) = {\rm Const.}  +  \alpha H,
\end{equation} 
and the albedos of objects within a given family or population are similar, then it follows from eq.~\ref{absmag} that the differential size distribution 
can be described as $n(D)\propto D^{-q}$, with the size distribution index  
\begin{equation} 
\label{qalpha}
                             q=5\alpha + 1.
\end{equation} 
While the absolute magnitude distributions derived here often cannot be 
described by a single power-law, eq.~\ref{qalpha} is still useful for 
locally relating the slope of the $H$ distribution to the slope of implied
(differential) $D$ distribution. For example, a model based on an 
equilibrium cascade in self-similar collisions developed by Dohnanyi (1969)
predicts $q=3.5$ and $\alpha=0.5$. 

\subsection{ The Comparison of Size Distributions for Families and Background }

\begin{figure*}[t]
\centering
\includegraphics[width=16cm]{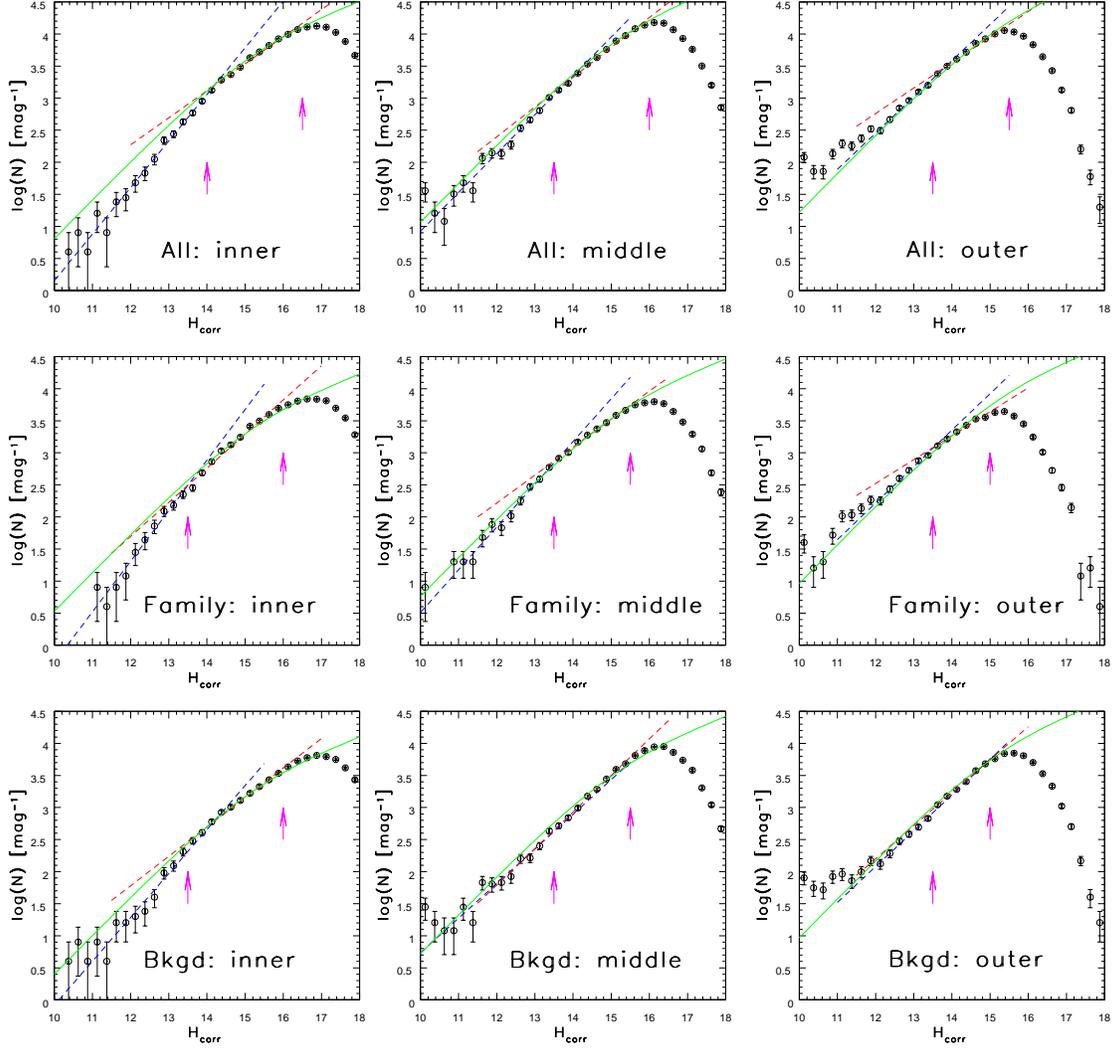}
%% BB 7 147 763 837
\caption{%\scriptsize 
The differential absolute magnitude distributions corresponding to panels in
Figure~\ref{prettyPlot4} are shown as symbols with (Poisson) error bars.
The solid line shows arbitrarily renormalized best-fit distribution from 
I01. The two dashed lines show the best-fit ``broken'' power law: a separate 
power-law fit for the bright and faint end. In some cases, the two lines are 
indistinguishable. The best-fit parameters are listed in Table 3. The two 
arrows show the best-fit break magnitude (left) and the adopted completeness 
limit (right).}
\label{9panelSD}
\end{figure*}

We compare the size distributions for the family population and for the
background, separately for the three regions defined by semi-major axis. 
The differential absolute magnitude distributions are shown in 
Figure~\ref{9panelSD}. To aid the comparison of different panels, we plot 
for reference the differential distribution derived from the {\it cumulative} 
distribution reported by I01
\begin{equation}
          n(r) = n_o {10^{ax} \over 10^{bx} + 10^{-bx}},
\end{equation}
where $x = H_{corr} - H_C$, $a = (k_1+k_2)/2$, $b = (k_1-k_2)/2$, with 
$H_C=15.5$, $k_1=0.65$ and $k_2=0.25$ (Table 4 in I01). 

I01 were able to fit this functional form because their sample extended
to a $\sim$1.5 mag fainter $H$ limit ($H_{corr}\sim 17.5$) than the sample 
discussed here. Given this sample difference, for each $H$ distribution 
shown in Figure~\ref{9panelSD} we instead fit a ``broken'' power law: 
a separate power-law fit for the bright and faint end. While this
procedure is expected to yield a shallower slope at the faint end than
the above I01 fit, it is preferred here because it ``decouples'' the bright
and faint ends. The separation of the bright and faint ends was 
attempted in $H$ steps of 0.5 mag, and the value that minimizes the resulting 
$\chi^2$ was adopted as the best fit. The statistical errors for the best-fit
slopes are typically 0.01-0.02, but it is likely that their uncertainty
is perhaps a factor of two or so larger due to systematic effects (see \S 
\ref{syserr} below). In a few cases, the best fit is consistent
with a single power law. The best-fit power-law parameters for differential 
absolute magnitude distributions shown in Figure~\ref{9panelSD} are 
listed in Table 3. 

The data and best fits shown in the top three panels in Figure~\ref{9panelSD}
demonstrate that the absolute magnitude distributions are not identical:
the outer main-belt shows a flatter distribution, and the inner belt shows
a steeper distribution than the middle belt region for objects with 
$H_{corr}<14$. This is in conflict with the I01 finding that the size 
distribution appears universal throughout the belt. However, here we analyze a sample 
about seven times larger; the statistical errors at the bright end
for the I01 sample were too large to detect this effect (see their Figures
21 and 22). Nevertheless, the I01 size distribution remains valid when the whole belt 
is treated together because the counts underprediction of their fit in the
outer belt is compensated by its overprediction in the inner belt. 

The separation of populations into families and background (the middle and
bottom rows in Figure~\ref{9panelSD}) shows that the flattening of $H_{corr}$
distribution as the semi-major axis increases is valid for each subpopulation
separately. Objects associated with families always show the flattening at
the faint end, while the background populations admit a single power-law
fit in the middle and outer belt. 

\begin{figure}[H]
\centering
\includegraphics[width=12cm]{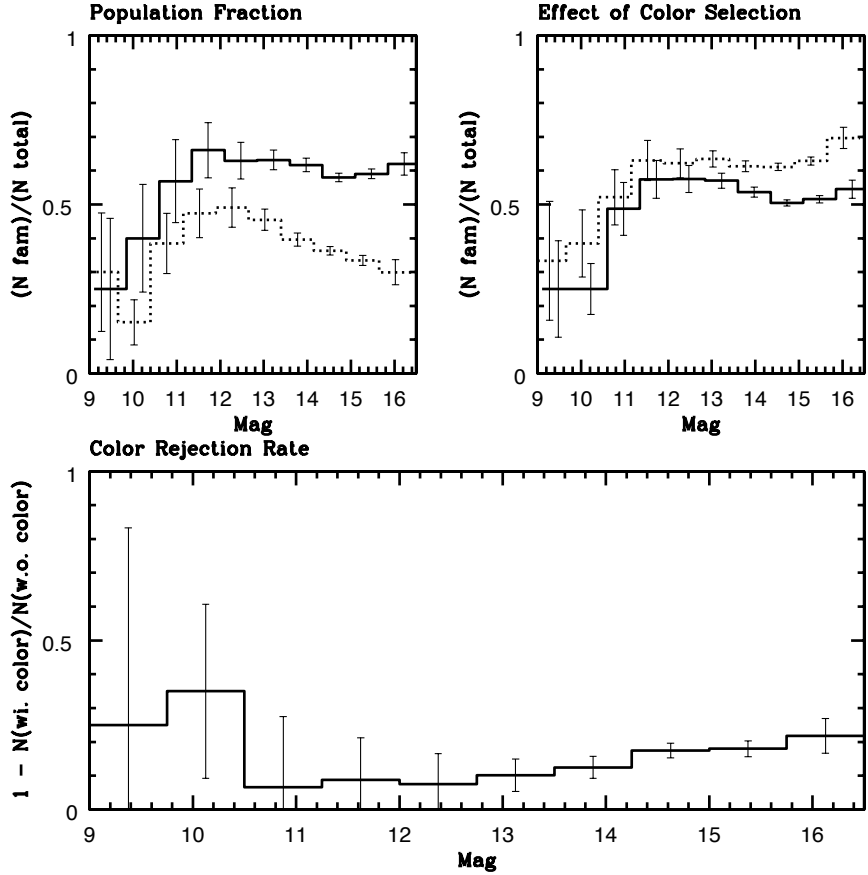}
\caption{
The family-to-background population ratios and effects of color selection 
on family populations. The top left panel shows the fraction of objects in 
families to the total population as a function of $H_{corr}$ magnitude, with 
the solid histogram representing red objects $(a^\ast > 0)$ and the dotted 
histogram representing blue objects $(a^\ast< 0)$. The top right panel
compares the fraction of objects in families to the total population as a 
function of magnitude when colors are used as a constraint on family membership 
(solid histogram) and when they are not (dotted histogram). The
bottom panel shows the rejected fraction due to color constraints as a 
function of magnitude.}
\label{Ratios}
\end{figure}

\begin{figure}[H]
\centering
\includegraphics[width=10cm]{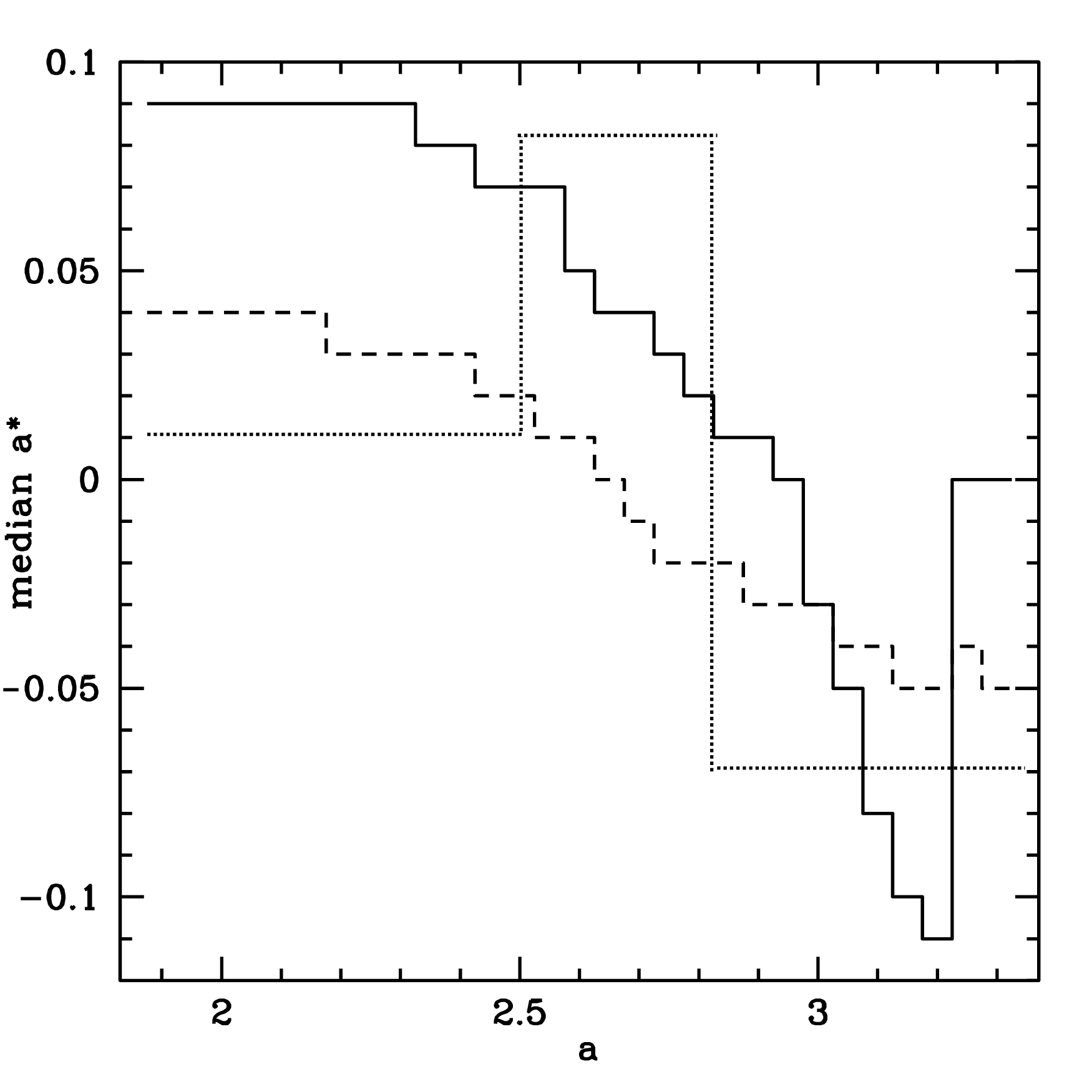}
\caption{Median $a^{*}$ color as a function of semi-major axis for several populations. Solid histogram represents the family population, dashed line represents the background population, and the dotted line represents the median of the family $a^\ast$ color centroids (unweighted by family size) for each Main Belt region (inner, mid, and outer). }
\label{avgCol}
\end{figure}

Due to different $H_{corr}$ distributions for family and background 
populations, the fraction of objects associated with families is a
function of $H_{corr}$. The top left panel in Figure~\ref{Ratios} 
shows this dependence separately for blue (dominated by the C 
taxonomic type) and red (dominated by the S type) subsets. For
both subsets, the fraction of objects in families significantly 
increases from $\sim$20\% to $\sim$50\% between $H_{corr}=9$ and 
$H_{corr}=11$. The two color-selected subsets show different 
behavior for $H_{corr}>11$: for blue subset the fraction of objects
in families decreases from $\sim$50\% to $\sim$30\%, while it 
stays constant at the $\sim$60\% level for red families. 
Since blue families typically have larger semi-major axis than
red families, it is possible that this decrease in family membership
is due to increasing color rejection at the faint end.
 However, the remaining two panels in Figure~\ref{Ratios} demonstrate that 
this is not the case because the color rejection rate is both fairly 
independent of $H_{corr}$, and too small to account for the observed
decrease of blue family membership. 

The dominance of background objects for $H_{corr} < 13$ is consistent
with the background population having a significantly shallower size distribution for large objects than the
families. The falloff of blue family fraction and the slow climb in color rejection
rate toward large values of $H_{corr}$ in Figure~\ref{Ratios} confirms that the size distributions for blue families 
are shallower for values of $H_{corr} > 13$. Because the red family fraction is effectively flat to our detection limit
it appears that the red family and background populations have identical size distributions for objects with $H_{corr} > 13$.
Morbidelli \emph{et al} (2003) suggest that the background
population is composed of many small families which formed from small-diameter objects and 
as such should have steep size distributions for objects larger than 1 km, producing a background population
with an initial size distribution (for objects $>$1 km) steeper than that for families which formed from the breakup of larger objects.
Differences between the size distributions of these two populations should eventually disappear through collisional processing.
Because the blue family and background populations appear to have significantly different size distributions  while the
red populations' size distributions appear identical (for $H_{corr} > 13$), we infer that the time for equilibrating the background and family size distributions
is longer for blue objects than red. This difference in the equilibration time may be due either to differences in asteroid internal structure and material properties between
taxonomic classes or due to environmental variations in collisional processing rates, as red objects are more prevalent in the inner belt and blue objects in the outer belt.

We note that the background population in the outer belt shows a curious
excess of large objects ($H_{corr}<11.5$) compared to best power-law fit (Figure~\ref{9panelSD}, bottom right). We 
have inspected the orbital parameter and color distributions for 58 objects
with $10<H_{corr}<11$ and found that they are not associated with any
identified family, nor generally clustered.

\subsection{ The Comparison of Size Distributions for Individual Families }

\begin{figure*}[H]
\centering
\includegraphics[width=16cm]{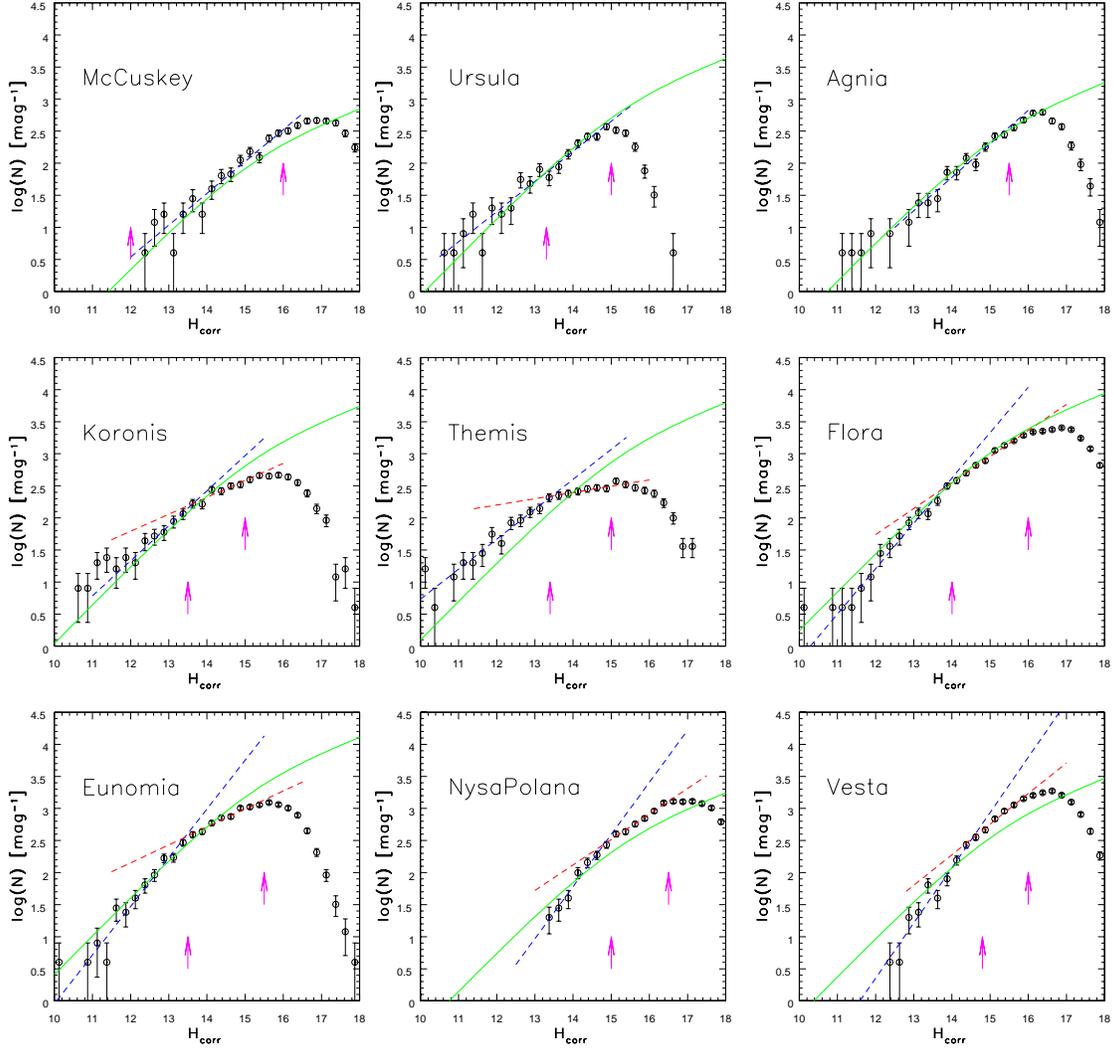}
%% BB 7 147 763 837
\caption{
Analogous to Figure~\ref{9panelSD}, except that the absolute magnitude 
distributions for selected asteroid families are shown. The first
three panels (from top left to bottom right) show examples of families that 
follow a single power-law magnitude distribution, and the remaining six panels
show magnitude distributions for families that require a double power-law
fit. The best-fit parameters are listed in Tables 4 and 5, respectively.}
\label{9panelSDfam}
\end{figure*}

The inspection of differential $H_{corr}$ distributions for the 37 families
identified here shows that many, but not all, display a clear change of slope 
such as seen for family populations in Figure~\ref{9panelSD}. We have
attempted a ``broken'' power-law fit for all families. When the two best-fit
slopes differ by less than 0.05, we enforce a single power-law fit. This
procedure yields 22 families described by a single power law and 15 families
with a robust detection of the slope change. Their best-fit parameters are
listed in Tables 4 and 5, respectively, and a few examples of measured $H$ 
distributions and best fits are shown in Figure~\ref{9panelSDfam}.  

For families whose absolute magnitude distributions are described by a single 
power law, the median best-fit power-law slope is 0.56, with a standard 
deviation of 0.16 (determined form inter-quartile range). This scatter is 
significantly larger than the measurement errors and indicate that 
{\it families do not have a universal size distribution}. Similarly, for 
families with a best-fit ``broken'' power law, the medians and standard
deviations for the ``bright'' and ``faint'' slopes are (0.66, 0.24)
and (0.32, 0.15), respectively (again note the significant scatter relative to the 
measurement errors), with the median $H_{corr}$ where the
slope changes of 14.2 ($D\sim$6 km for $p_V=0.1$). We discuss correlations of these best-fit 
parameters with the family color and age in \S \ref{corr}.

\subsection{ Systematic Deviations in Size Distribution due to Variations 
                                         in  Family Definitions  } 

\label{syserr}

Before proceeding with the analysis of correlations between size distributions
and other family properties such as color and age, we analyze the systematic 
deviations in size distribution due to variations in family definitions. 
For example, the color constraints may result in a size-dependent
incompleteness because of the increased photometric noise at the faint end.
Similarly, the assumption of gaussian distributions for orbital parameters
and colors may result in incomplete families due to extended halos, as 
pointed out by N05. This effect may also induce size-dependant systematics
because small objects are scattered over a larger region of orbital space, as shown below.

\begin{figure}[H]
\centering
\includegraphics[width=7cm]{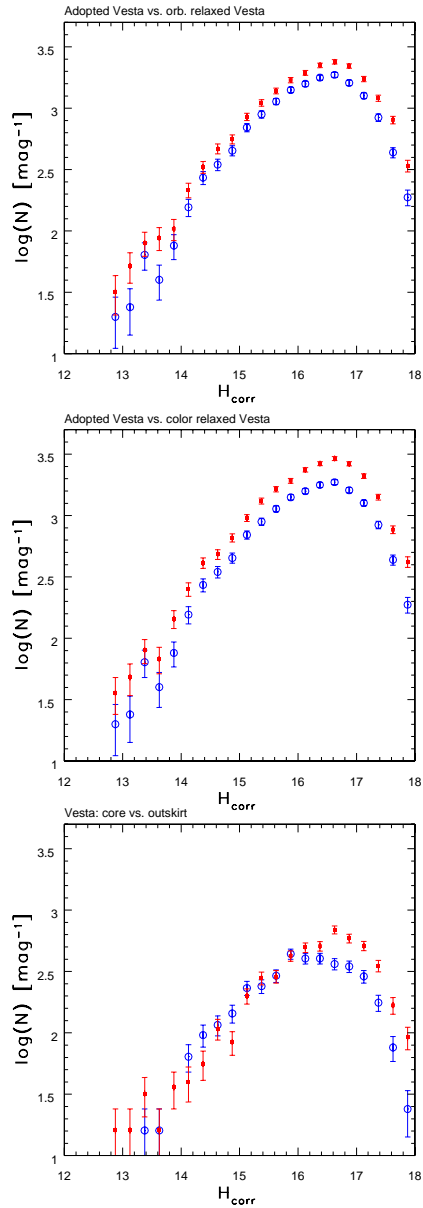}
%% BB 151 150 439 830
\caption{
The dependence of absolute magnitude distribution on family definition
for Vesta family. The top panel compares the magnitude distribution 
obtained for adopted family definition (open circles) to that obtained
for a much less constraining cut on orbital elements (closed squares). The middle 
panel compares the adopted distribution (open circles) to that obtained
for a much less constraining color cut (closed squares). In both cases the
shapes of the magnitude distributions are similar, with only significant
change in the number of selected candidate members. The bottom panel
separates the adopted population into the orbital ``core'' region (open circles)
and ``outskirt'' region (closed squares). Note that the latter distribution
is steeper (i.e. the small members are more prevalent in the ``outskirt''
region).}
\label{vesta}
\end{figure}

The Vesta family offers a good test case because of its unique color 
distribution (which is due to the influence of 1 $\mu$m absorption feature
on the measured $i-z$ color). The top panel in Figure~\ref{vesta} compares 
the $H_{corr}$ distributions for the adopted Vesta family and for a less 
constraining orbital cut defined simply by $0.06 < sin(i) < 0.16$ and 
$e<0.16$, that yields 30\% more candidate members. Apart from this
overall shift in the normalization, the resulting distributions have
statistically indistinguishable shapes. The middle panel compares the
adopted family and a much less constraining color cut, $a^\ast>0$ (i.e.
no constraint on the $i-z$ color), that yields 50\% more objects. 
Again, the slope of the two distributions are indistinguishable. 

We detect a significant difference, however, when we split the adopted
family in the ``core'' and ``outskirt'' parts using $D_{orbit}<1$ and
$1.75 < D_{orbit} < 2.75$ (see eq.~\ref{Dorb}). As the bottom panel
in Figure~\ref{vesta} shows, the ``outskirt'' subsample has a steeper
$H_{corr}$ distribution than the ``core'' subsample. The best-fit
power-law slopes in the $14 < H_{corr} < 16$ region are 0.45 and 0.59
for the ``core'' and ``outskirt'' subsample, respectively. We note
that despite this slope difference, the change of slope between the
bright (0.89) and faint end is robustly detected. 

Another method to see the same ``size sorting'' effect is to inspect the
dependence of $D_{orbit}$ on $H_{corr}$. We find that the median 
$D_{orbit}$ for objects in the Vesta family increases from 1.0 to 1.5 as 
$H_{corr}$ increases from 14 to 17. 

\begin{figure}[H]
\centering
\includegraphics[width=12cm]{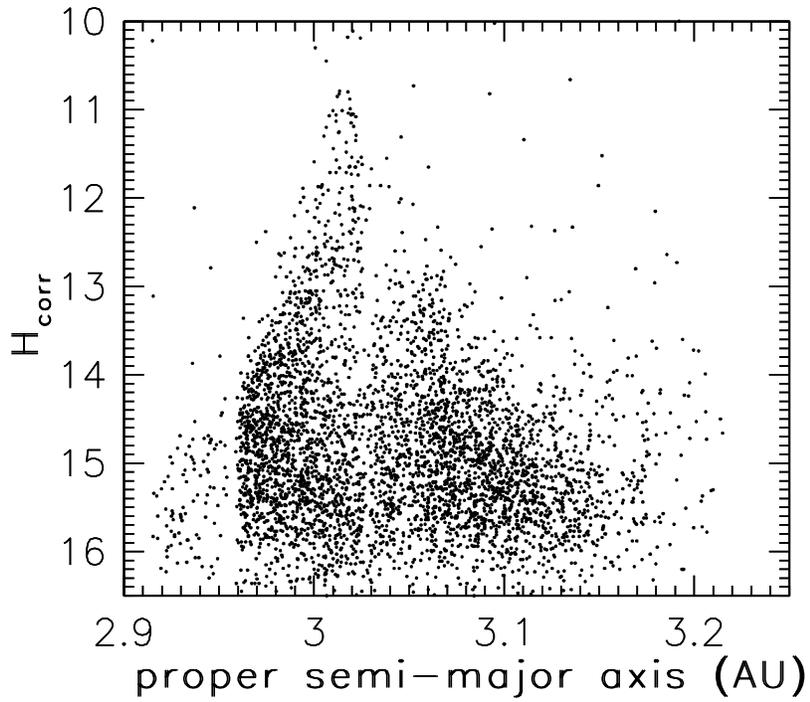}
% BB from 20 17 592 779 to 30 317 582 729
\caption{
The correlation between the proper semi-major axis and absolute magnitude
for the Eos family. The increased dispersion of the semi-major axis for
faint (small) objects is probably caused by the Yarkovsky effect. Note that
this family is intersected by several mean motion resonances with Jupiter
(e.g. 7:3 at $a_p \sim 2.95$).}
\label{Yark}
\end{figure}

This ``size sorting'' effect is not a peculiar property of the
Vesta family as it is seen for a large fraction of families.
It is caused by an increased scatter in all three orbital
parameters as $H_{corr}$ increases. This is not surprising as the
velocity field of the fragments produced in the disruption of an asteroid family's
parent body may have been size-dependent. For most families the sorting is 
dominated by the increased dispersion in the semi-major axis.
One of the most striking examples, the Eos family, is 
shown in Figure~\ref{Yark}. As discussed by N05, this increase of
dispersion as size decreases can be also be explained as the drift induced by the
Yarkovsky effect (see also Vokrouhlick\'{y} 1999; and Bottke et al. 2001).

\subsection{ Correlations between Size Distributions and Family Color and Age }
\label{corr}

We analyze the correlations between the best-fit size 
distribution parameters listed in Tables 4 and 5, and 
family color and age. The age, when available, is taken
from the compilation by N05. 

\begin{figure*}[H]
\centering
\includegraphics[width=12cm]{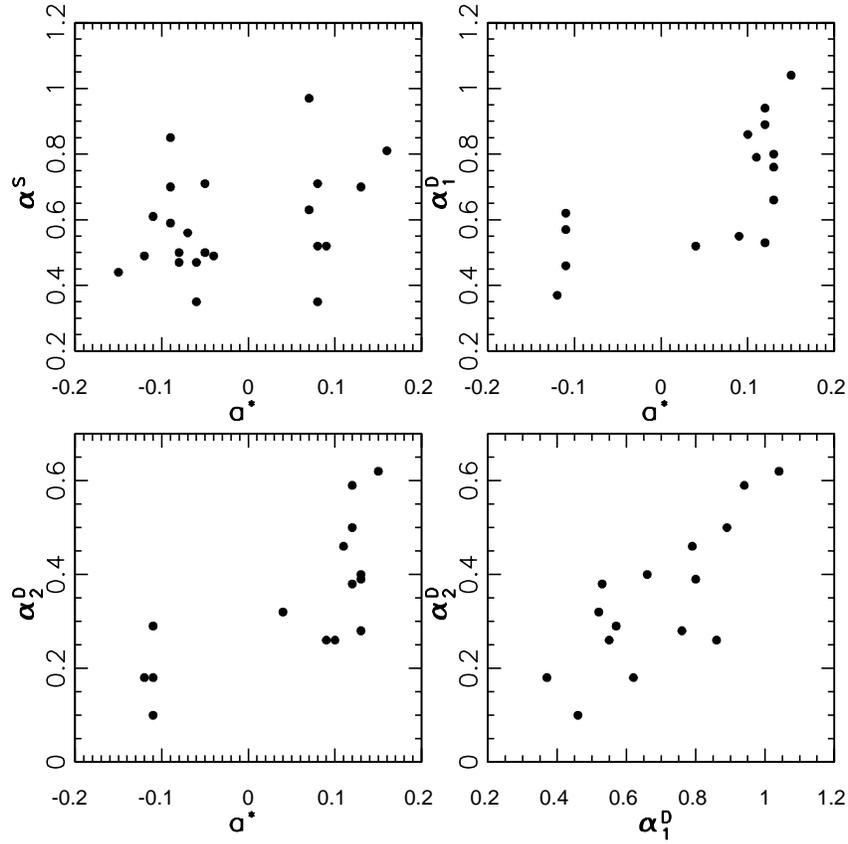}
%% BB  20 217 592 779
\caption{
A summary of relationships between the family median color and 
parameters of the  best-fit power-law magnitude distributions.
The top left panel shows the slope of the  best-fit power-law 
as a function of color $a^*$ (see text for definition) for 25
families that follow a single power-law magnitude distribution.
The top right panel shows the ``bright'' slope of the best-fit
double power-law as a function of color $a^*$ for 12 families 
that follow a double power-law magnitude distribution, and
the bottom left panel is analogous plot for the ``faint'' slope.
Note that blue (C type) families have much shallower magnitude
distributions than redder families. The bottom right panel 
demonstrates the strong correlation between the ``bright'' 
and ``faint'' slopes, that seems independent of color. 
The errors of alpha are dominated by systematics. 
As per discussion in Section 4.3, we estimate these
systematic errors to be about 0.05-0.1.}
\label{slopes}
\end{figure*}

The dependence of the power-law index on the mean $a^\ast$ color
for families described by a single power law is shown in the
top left panel in Figure~\ref{slopes}. The mean and standard 
deviation for 14 blue families are (0.55, 0.13), and for 8 red 
families are (0.65, 0.19). These differences are not statistically
significant. Within each color-selected subsample (blue vs. red, i.e. $a^{*} < 0$ vs $a^{*} > 0$),
there is no discernible correlation between the slope and color. 

Families that require a ``broken'' power law fit are twice as likely 
for red families ($a^\ast>0$) dominated by S type asteroids) than
for blue families dominated by C type asteroids (73\% vs. 36\%). 
As illustrated in Figure~\ref{slopes}, the size distributions are 
systematically steeper for S type families, and the ``bright'' and
``faint'' end slopes appear to be correlated. The median values 
of the ``bright'' and ``faint'' end slopes are (0.57, 0.18) for blue 
families, and (0.79, 0.39) for red families.

For a subset of families that have available age estimates, 
we find that families with ``broken'' power law size distributions
are dominated by old families, while those that admit a single
power law are dominated by young families, with the age separation
boundary at $\sim$1 Gyr. We note that the size distribution was
used for some of age estimates compiled by N05, so this conclusion
may be a bit of circular reasoning, though the majority of age estimates 
are derived independently of the observed size distribution.

\begin{figure}[H]
\centering
\includegraphics[width=12cm]{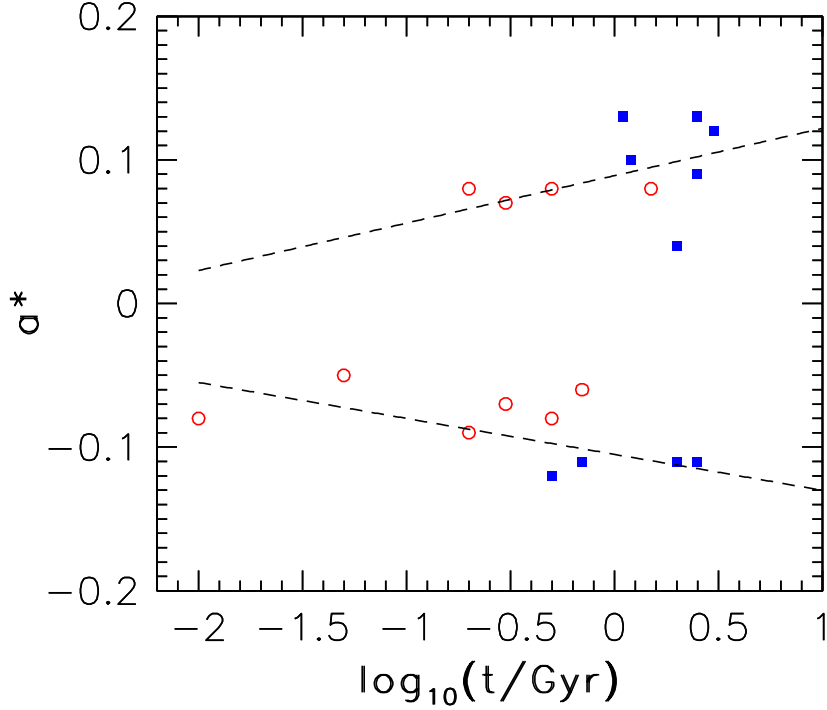}
% BB from 20 17 592 779 to 30 317 582 729
\caption{
The correlation between the mean $a^\ast$ color for families defined
here and their age taken from N05. Familes whose absolute magnitude
distribution can be described by a single power law are shown 
by circles, and those that require a broken power law as squares.
Note that the former are dominated by young families ($\leq$1 Gyr),
and the latter by old families.
The two lines are the best-fit color-age relation from N05, 
converted using $a^\ast = 0.49\,PC_1 -0.16$, where $PC_1$ is the
first principal color component derived by N05 (note that 
N05 used several very young familes not discussed here to 
constrain the slopes of plotted relations).} 
\label{colorAge}
\end{figure}

The correlations between the mean color and family age
reported by Jedicke et al. (2004) and N05 are reproduced
when using the $a^\ast$ color for families discussed here.
Figure~\ref{colorAge} illustrates a good agreement with
the analytic fits to the observed correlations obtained 
by N05, and further demonstrates the correlation between
the observed size distributions and age. 

\section{    Discussion and Conclusions   }

We have used a large sample of asteroids ($\sim$88,000) for which 
both orbital elements and SDSS colors are available to derive 
improved membership for 37 asteroid families. The addition of colors 
typically rejects about 10\% of all dynamically identified candidate members
due to mismatched colors. Four families can be reliably isolated only with 
the aid of colors. About 50\% of objects in this data set belong to families, 
with this fraction representing a lower limit due to a conservative 
requirement that a candidate family must include at least 100 members.
The resulting family definitions are in good agreement with previous
work (e.g. Z95, N05) and all the discrepancies are well understood. 
Although SDSS has observed only about 1/3 of all known asteroids, 
it is remarkable that the sample discussed here provides color
information for more than an order of magnitude more objects 
associated with families than analyzed in the published literature. 

This data set enables the determination of absolute magnitude (size) 
distributions for individual families to a very faint limit without a 
need to account for complex selection effects. We verify that size distribution 
varies significantly among families, and is typically different from size 
distributions for background populations. Consequently, the asteroid 
size distribution cannot be described by a universal function that is
valid throughout the main belt (e.g. Jedicke \& Metcalfe 1998, Ivezi\'{c}
et al. 2001, and reference therein). This finding will have an influence 
on conclusions derived from modeling the size distribution under this
assumption (e.g. Bottke et al. 2005, and references therein). In 
particular, it is not clear how to interpret a detailed dependence of
the critical specific energy (energy per unit mass required to fragment
an asteroid and disperse the fragments to infinity) on asteroid size
derived from such models, when the starting observational constraint 
on size distribution is an average over multiple families with significantly
varying size distributions. 

We show that for objects with $H_{corr} < 13$, the background population
dominates (family fraction decreases toward lower $H_{corr}$, indicating 
a shallower size distribution for large objects), 
while for objects with $H_{corr} > 13$, the red family fraction 
remains effectively constant to our completeness limit
 while the blue family fraction falls off. This indicates that the time 
 to collisionally equilibrate the family and background populations 
 (see e.g., Morbidelli et al. 2003)
 is shorter for red objects than blue.

The size distributions for 15 families display a well-defined change of slope 
and can be modeled as a ``broken'' double power-law. The first evidence for 
this effect and a discussion of its significance are presented by Morbidelli 
et al. (2003). Using a data set with much simpler correction for the
observational selection effects, we confirm their result in a statistically 
more robust way. We also find such ``broken'' size 
distributions are twice as likely for S-type familes than for C-type families 
(73\% vs. 36\%), and are dominated by dynamically old families. The remaining 
families with size distributions that can be modeled as a single power law 
are dominated by young families ($<$1 Gyr). 

The eight largest families all show a change of size distribution slope to 
much smaller values at the faint end (see Table 5). This result has a
direct consequence when prediciting the number of very small objects ($D\sim1$
km). In particular, it could explain why the Statistical Asteroid Model 
developed by Tedesco, Cellino \& Zappal\'{a} (2005) predicts too many 
objects: the data presented here are inconsistent with the SAM assumptions
for the number of objects in its most populous families such as Eunomia
and Themis. 

We find that when size distribution requires a double power-law model, the 
two slopes are correlated and are steeper for S-type families. No such 
slope--color correlation is discernible for families whose size distribution 
follows a single power law. While beyond the scope of this work, the modeling
of such correlations may shed light on the internal structure and material properties of asteroids. 

For several very populous families, we find that the size distribution varies 
with the distance from the core in orbital-color space, such that small
objects are more prevalent in the family outskirts. As discussed by N05 (and
references therein), this ``size sorting'' is consistent with predictions
based on the Yarkovsky/YORP effect. 

While these results provide significant new observational constraints
for the properties of main-belt asteroids, very soon the observations
will further improve. The upcoming large-scale sky surveys, such as Pan-STARRS 
(Kaiser et al. 2002) and LSST (Tyson 2002), will obtain even more impressive
samples, both in size, diversity of measurements and their accuracy. 
For example, LSST will scan the whole observable sky every three nights 
in two bands to a $5\sigma$ depth equivalent to $V=24.7$. These data will
enable much improved analysis due to several factors
\begin{itemize}
\item Due to hundreds of observations, the orbits will be determined
directly, instead of relying on external data, resulting in a sample
about 30-40 times larger than discussed here
\item The effective faint limit will be extended by about 5 magnitudes,
correspoding to ten times smaller size limit (diameters of several hundred 
meters) 
\item Due to many photometric observations obtained with the same 
well-calibrated system, the uncertainties in absolute magnitudes will
be an order of magnitude smaller
\item The addition of the $y$ band (at $\sim$1 $\mu$m) will improve
the color classification due to better sensitivity to the $\sim$1 $\mu$m
absorption feature present in spectra of many asteroids. 
\end{itemize}
These new data will undoubtely reinvigorate both observational and theoretical 
studies of main-belt asteroids.

\section{Acknowledgments}
We are grateful to E. Bowell for making his ASTORB file publicly available, 
and to A. Milani, Z. Kne\v{z}evi\'{c} and their collaborators for generating 
and distributing proper orbital elements. M.J. gratefully acknowledges 
support from the Taplin Fellowship and from NSF grant PHY-0503584.

Funding for the SDSS and SDSS-II has been provided by the Alfred P. Sloan 
Foundation, the Participating Institutions, the National Science Foundation, 
the U.S. Department of Energy, the National Aeronautics and Space 
Administration, the Japanese Monbukagakusho, the Max Planck Society, and 
the Higher Education Funding Council for England. The SDSS Web Site is 
http://www.sdss.org/.

The SDSS is managed by the Astrophysical Research Consortium for the
Participating Institutions. The Participating Institutions are the American
Museum of Natural History, Astrophysical Institute Potsdam, University of
Basel, University of Cambridge, Case Western Reserve University, University of
Chicago, Drexel University, Fermilab, the Institute for Advanced Study, the
Japan Participation Group, Johns Hopkins University, the Joint Institute for
Nuclear Astrophysics, the Kavli Institute for Particle Astrophysics and
Cosmology, the Korean Scientist Group, the Chinese Academy of Sciences
(LAMOST), Los Alamos National Laboratory, the Max-Planck-Institute for
Astronomy (MPIA), the Max-Planck-Institute for Astrophysics (MPA), New Mexico
State University, Ohio State University, University of Pittsburgh, University
of Portsmouth, Princeton University, the United States Naval
Observatory, and the University of Washington.

\newcommand{\aj}{Astron. J.}
\newcommand{\apj}{Astrophys. J.}
\newcommand{\apjs}{Astrophys. J. Suppl. Ser.}
\newcommand{\aap}{Astron. Astrophys.}
\newcommand{\mnras}{Mon. Not. R. Astron. Soc.}
\newcommand{\nat}{Nature}
\newcommand{\pasj}{Pub. Astron. Soc. Japan}
\newcommand{\jgr}{J. Geophys. Res.}

\nocite{*}
 \bibliographystyle{icarus}
\bibliography{paperV2.5_pre.bib}

\clearpage
\pagebreak

%%%%FIGURES%%%%%%%

\clearpage
\pagebreak
%%%%%TABLES%%%%%%%

{
\renewcommand{\baselinestretch}{1}
\small\normalsize

\textwidth= 7.5in
\oddsidemargin= -0.5in

\begin{table}
\scriptsize
\centering
\begin{tabular}{|r|c|c|r|r|c|c|r|c|r|c|c|c|c|c|}
\hline
Family$^a$  & 
$a_p^b$ & $\sigma_a^b$ &
$i_p^c$ & $\sigma_i^c$ &
$e_p^d$ & $\sigma_e^d$ &
$a^e$   & $\sigma_a^e$ &
$i-z^f$   & $\sigma_{iz}^f$ &
$\Delta_{Orb}^g$ & $\Delta_{Col}^h$ &
$a_{min}^i$ & $a_{max}^i$ \\
\hline
\hline
Flora$^{1}$ & 2.272 & 0.065 & 0.084 & 0.023 & 0.133 & 0.027 & 0.130 & 0.044 & $-$0.040 & 0.059 & 2.60 & 2.75 & 1.50 & 2.50  \\
Baptistina & 2.277 & 0.035 & 0.096 & 0.006 & 0.143 & 0.006 & $-$0.050 & 0.030 & 0.050 & 0.067 & 2.20 & 3.00 & 1.50 & 2.50  \\
Vesta$^{1}$ & 2.350 & 0.072 & 0.114 & 0.009 & 0.100 & 0.015 & 0.120 & 0.052 & $-$0.270 & 0.089 & 2.75 & 3.50 & 1.50 & 2.50  \\
Erigone$^{1}$ & 2.371 & 0.028 & 0.087 & 0.006 & 0.207 & 0.007 & $-$0.090 & 0.037 & 0.060 & 0.082 & 2.50 & 3.10 & 1.50 & 2.50  \\
McCuskey & 2.374 & 0.065 & 0.049 & 0.008 & 0.151 & 0.015 & $-$0.120 & 0.052 & 0.010 & 0.074 & 2.50 & 2.20 & 1.50 & 2.50  \\
Euterpe & 2.374 & 0.076 & 0.017 & 0.007 & 0.185 & 0.011 & 0.100 & 0.052 & $-$0.040 & 0.104 & 2.00 & 3.00 & 1.50 & 2.50  \\
Nysa-Polana$^{1}$ & 2.388 & 0.044 & 0.042 & 0.005 & 0.184 & 0.019 & 0.130 & 0.052 & $-$0.040 & 0.067 & 3.00 & 3.50 & 1.50 & 2.50  \\
Massalia$^{1}$ & 2.405 & 0.029 & 0.025 & 0.002 & 0.163 & 0.006 & 0.070 & 0.052 & $-$0.040 & 0.082 & 3.50 & 2.50 & 1.50 & 2.50  \\
Andree & 2.405 & 0.034 & 0.084 & 0.005 & 0.170 & 0.013 & 0.140 & 0.052 & $-$0.030 & 0.089 & 3.00 & 3.00 & 1.80 & 2.50  \\
Teutonia & 2.574 & 0.039 & 0.070 & 0.021 & 0.169 & 0.049 & 0.120 & 0.052 & $-$0.030 & 0.089 & 2.20 & 2.60 & 2.50 & 2.82  \\
Rafita$^{1}$ & 2.584 & 0.024 & 0.131 & 0.004 & 0.173 & 0.005 & 0.080 & 0.037 & $-$0.040 & 0.067 & 2.50 & 3.00 & 2.50 & 2.82  \\
Maria$^{1}$ & 2.612 & 0.053 & 0.254 & 0.008 & 0.090 & 0.021 & 0.120 & 0.037 & $-$0.010 & 0.059 & 2.50 & 3.00 & 2.50 & 2.82  \\
Mitidika & 2.618 & 0.067 & 0.216 & 0.013 & 0.244 & 0.023 & $-$0.110 & 0.044 & 0.030 & 0.067 & 2.50 & 3.00 & 2.50 & 2.82  \\
Eunomia$^{1}$ & 2.632 & 0.073 & 0.227 & 0.011 & 0.150 & 0.018 & 0.130 & 0.044 & $-$0.020 & 0.052 & 2.50 & 3.00 & 2.50 & 2.82  \\
Misa$^{1}$ & 2.650 & 0.035 & 0.040 & 0.004 & 0.178 & 0.007 & $-$0.080 & 0.052 & 0.050 & 0.067 & 3.00 & 3.00 & 2.50 & 2.82  \\
Adeona$^{1}$ & 2.660 & 0.038 & 0.202 & 0.004 & 0.168 & 0.007 & $-$0.110 & 0.037 & 0.060 & 0.059 & 2.60 & 2.80 & 2.50 & 2.82  \\
Juno$^{1}$ & 2.664 & 0.026 & 0.230 & 0.004 & 0.234 & 0.005 & 0.080 & 0.052 & $-$0.030 & 0.082 & 3.00 & 3.00 & 2.50 & 2.82  \\
Aeolea$^{1}$ & 2.671 & 0.033 & 0.066 & 0.004 & 0.190 & 0.005 & $-$0.110 & 0.044 & 0.040 & 0.096 & 4.00 & 3.00 & 2.50 & 2.82  \\
Henan & 2.694 & 0.069 & 0.045 & 0.011 & 0.057 & 0.013 & 0.110 & 0.044 & $-$0.010 & 0.111 & 2.00 & 2.60 & 2.50 & 2.82  \\
Nemesis$^{1}$ & 2.733 & 0.018 & 0.085 & 0.003 & 0.087 & 0.003 & $-$0.080 & 0.037 & 0.040 & 0.089 & 2.50 & 2.50 & 2.50 & 2.82  \\
Lydia & 2.737 & 0.048 & 0.105 & 0.005 & 0.029 & 0.019 & 0.160 & 0.052 & 0.010 & 0.096 & 2.50 & 2.50 & 2.50 & 2.82  \\
Padua$^{1}$ & 2.740 & 0.035 & 0.090 & 0.005 & 0.042 & 0.011 & $-$0.050 & 0.044 & 0.030 & 0.082 & 2.70 & 3.50 & 2.50 & 2.82  \\
Chloris$^{1}$ & 2.744 & 0.040 & 0.152 & 0.005 & 0.254 & 0.008 & $-$0.060 & 0.030 & 0.050 & 0.052 & 3.00 & 3.00 & 2.50 & 2.82  \\
Merxia$^{1}$ & 2.751 & 0.036 & 0.087 & 0.002 & 0.136 & 0.006 & 0.090 & 0.037 & $-$0.070 & 0.096 & 3.00 & 3.00 & 2.50 & 2.82  \\
Gefion$^{1}$ & 2.768 & 0.029 & 0.159 & 0.004 & 0.134 & 0.020 & 0.100 & 0.044 & $-$0.020 & 0.067 & 3.00 & 3.00 & 2.50 & 2.82  \\
Agnia$^{1}$ & 2.761 & 0.046 & 0.063 & 0.013 & 0.070 & 0.012 & 0.090 & 0.059 & 0.000 & 0.104 & 2.20 & 2.60 & 2.50 & 2.82  \\
Dora$^{1}$ & 2.787 & 0.019 & 0.136 & 0.002 & 0.195 & 0.003 & $-$0.120 & 0.037 & 0.050 & 0.052 & 3.50 & 2.80 & 2.50 & 2.82  \\
Brasilia$^{1}$ & 2.851 & 0.009 & 0.259 & 0.002 & 0.123 & 0.002 & $-$0.050 & 0.037 & 0.050 & 0.059 & 3.50 & 3.20 & 2.82 & 3.05  \\
Koronis$^{1}$ & 2.904 & 0.046 & 0.037 & 0.002 & 0.053 & 0.019 & 0.090 & 0.044 & $-$0.020 & 0.059 & 3.00 & 3.00 & 2.82 & 3.05  \\
Eos$^{1}$ & 3.021 & 0.059 & 0.175 & 0.012 & 0.073 & 0.015 & 0.050 & 0.044 & 0.030 & 0.052 & 3.50 & 3.00 & 2.91 & 3.60  \\
Tirela$^{1}$ & 3.122 & 0.032 & 0.285 & 0.007 & 0.195 & 0.016 & 0.150 & 0.052 & 0.080 & 0.067 & 3.00 & 2.50 & 2.91 & 3.60  \\
Themis$^{1}$ & 3.126 & 0.071 & 0.024 & 0.008 & 0.153 & 0.019 & $-$0.120 & 0.037 & 0.010 & 0.059 & 2.50 & 2.50 & 2.91 & 3.60  \\
Hygiea$^{1}$ & 3.144 & 0.066 & 0.090 & 0.008 & 0.137 & 0.019 & $-$0.110 & 0.037 & 0.010 & 0.082 & 2.50 & 2.50 & 2.91 & 3.60  \\
Lixiaohua$^{1}$ & 3.149 & 0.015 & 0.177 & 0.004 & 0.197 & 0.005 & $-$0.070 & 0.037 & 0.080 & 0.067 & 3.00 & 2.50 & 2.91 & 3.60  \\
Ursula & 3.156 & 0.060 & 0.279 & 0.012 & 0.090 & 0.028 & $-$0.070 & 0.044 & 0.060 & 0.067 & 2.50 & 3.00 & 2.91 & 3.60  \\
Veritas$^{1}$ & 3.168 & 0.007 & 0.160 & 0.003 & 0.063 & 0.004 & $-$0.080 & 0.030 & 0.050 & 0.052 & 3.00 & 3.50 & 2.91 & 3.60  \\
Theobalda & 3.170 & 0.014 & 0.247 & 0.006 & 0.251 & 0.011 & $-$0.160 & 0.037 & 0.040 & 0.089 & 2.75 & 3.50 & 3.00 & 3.50  \\
\hline
\end{tabular}
%\end{center}
\\
\vspace{1mm}
{
 \scriptsize
 {\it Notes:}\\
$^a$ the family name (the lowest-numbered member), sorted by semi-major axis \\
$^b$ the mean proper semi-major axis (AU) and its gaussian dispersion
     ($\sigma$) adopted for this family  \\
$^c$ the mean sine of proper inclination  and its gaussian dispersion
     adopted for this family \\
$^d$ the mean proper orbital eccentricity and its gaussian dispersion
     adopted for this family \\
$^e$ the mean $a^\ast$ color (see text for definition) and its gaussian dispersion
     adopted for this family \\
$^f$ the mean $i-z$ color and its gaussian dispersion
     adopted for this family \\
$^g$ the maximum deviation in orbital space from the adopted mean values,
     in units of adopted dispersions (see text) \\
$^h$ the maximum deviation in color space from the adopted mean values,
     in units of adopted dispersions (see text) \\
$^i$ the minimum and maximum semi-major axis adopted for this family \\
$^{1}$ family matched to N05 \\
}
\caption{Adopted family definitions in the orbital and color space. }
\end{table}
%%%%%%%%%%%%%%%%%%%%%%%%

\begin{table}
\scriptsize
\centering
\begin{tabular}{|r|r|r|r|r|r|r|r|r|}
\hline
Family$^a$  & $N^b$ &
$a_p^c$ & $i_p^d$ & $e_p^e$ & 
 a$^f$ &  $i-z^g$ \\ 
\hline
\hline
Flora &  6164 &  2.28 &  0.08 &  0.13 &  0.13 & $-$0.05  \\
Baptistina &   310 &  2.28 &  0.10 &  0.14 & $-$0.04 &  0.03  \\
Vesta &  3793 &  2.35 &  0.11 &  0.10 &  0.12 & $-$0.32  \\
McCuskey &  1043 &  2.36 &  0.05 &  0.15 & $-$0.12 &  0.01  \\
Erigone &   307 &  2.37 &  0.09 &  0.21 & $-$0.09 &  0.05  \\
Euterpe &   387 &  2.38 &  0.02 &  0.18 &  0.09 & $-$0.03  \\
Nysa-Polana &  2928 &  2.39 &  0.04 &  0.18 &  0.13 & $-$0.04  \\
Andree &   649 &  2.40 &  0.08 &  0.17 &  0.13 & $-$0.02  \\
Massalia &   730 &  2.41 &  0.03 &  0.16 &  0.07 & $-$0.04  \\
Teutonia &  3405 &  2.57 &  0.07 &  0.16 &  0.12 & $-$0.04  \\
Rafita &   225 &  2.59 &  0.13 &  0.17 &  0.08 & $-$0.04  \\
Maria &  1315 &  2.61 &  0.25 &  0.09 &  0.12 & $-$0.02  \\
Mitidika &   698 &  2.61 &  0.22 &  0.24 & $-$0.11 &  0.03  \\
Eunomia &  2995 &  2.63 &  0.23 &  0.15 &  0.13 & $-$0.03  \\
Misa &   185 &  2.65 &  0.04 &  0.18 & $-$0.08 &  0.05  \\
Adeona &   428 &  2.66 &  0.20 &  0.17 & $-$0.11 &  0.05  \\
Juno &   354 &  2.66 &  0.23 &  0.23 &  0.07 & $-$0.03  \\
Aeolea &   172 &  2.66 &  0.07 &  0.19 & $-$0.09 &  0.04  \\
Henan &   624 &  2.67 &  0.04 &  0.06 &  0.11 & $-$0.02  \\
Nemesis &   129 &  2.73 &  0.09 &  0.09 & $-$0.09 &  0.02  \\
Lydia &   598 &  2.74 &  0.11 &  0.03 &  0.16 &  0.01  \\
Padua &   442 &  2.75 &  0.09 &  0.04 & $-$0.05 &  0.05  \\
Merxia &   252 &  2.75 &  0.09 &  0.14 &  0.08 & $-$0.07  \\
Gefion &   914 &  2.76 &  0.16 &  0.13 &  0.10 & $-$0.03  \\
Chloris &   121 &  2.76 &  0.15 &  0.25 & $-$0.06 &  0.05  \\
Agnia &  1106 &  2.76 &  0.06 &  0.07 &  0.08 & $-$0.01  \\
Dora &   248 &  2.79 &  0.14 &  0.20 & $-$0.12 &  0.04  \\
Brasilia &   127 &  2.85 &  0.26 &  0.12 & $-$0.05 &  0.03  \\
Koronis &  1267 &  2.90 &  0.04 &  0.05 &  0.09 & $-$0.02  \\
Eos &  4367 &  3.04 &  0.18 &  0.07 &  0.04 &  0.03  \\
Tirela &   411 &  3.12 &  0.29 &  0.20 &  0.15 &  0.08  \\
Themis &  1073 &  3.13 &  0.02 &  0.15 & $-$0.11 &  0.01  \\
Hygiea &  1076 &  3.15 &  0.09 &  0.13 & $-$0.11 &  0.01  \\
Lixiaohua &   150 &  3.15 &  0.18 &  0.20 & $-$0.07 &  0.05  \\
Ursula &   644 &  3.15 &  0.28 &  0.09 & $-$0.06 &  0.06  \\
Veritas &   250 &  3.17 &  0.16 &  0.06 & $-$0.08 &  0.05  \\
Theobalda &   100 &  3.17 &  0.25 &  0.25 & $-$0.15 &  0.01  \\
\emph{NA}$^h$ &   \emph{90} &  2.87 &  0.11 &  0.05 & 0.17 &  0.04  \\
\emph{NA}$^h$ &   \emph{5} &  2.77 &  0.08 &  0.09 & $-$0.08 &  0.14  \\
\emph{NA}$^h$ &   \emph{46} &  2.78 &  0.06 &  0.07 & 0.00 &  $-$0.25   \\
\hline
\end{tabular}
\\
\vspace{1mm}
 {
 \scriptsize
 {\it Notes:}\\
$^a$ the family name (the lowest-numbered member) \\
$^b$ the number of objects in SDSS MOC 4 associated with this family \\
$^c$ the median proper semi-major axis (AU) \\
$^d$ the median sin of proper inclination \\
$^e$ the median proper orbital eccentricity \\
$^f$ the median $a^\ast$ color (see text for definition) \\
$^g$ the median $i-z$ color \\
$^h$ rejected clumps with less than 100 members\\
}
\caption{The median orbital parameters and SDSS colors for detected families and rejected clumps (last three).}
\end{table}
%%%%%%%%%%%%%%%%%%%%%%%%

\begin{table}
\scriptsize
\centering
\begin{tabular}{|r|r|c|c|c|c|c|}
\hline
$a$ range$^a$ &  N$^b$ & H$_{min}^c$ & H$_{max}^c$ & H$^d_B$ & $\alpha^e_1$ & $\alpha^f_2$ \\
\hline
\hline
  2.00 -- 2.50 &  30,702 &  11.0 & 16.5 & 14.0 & 0.76 & 0.46 \\
       family  &  16,309 &  11.0 & 16.0 & 13.5 & 0.79 & 0.53 \\
   background  &  14,393 &  11.0 & 16.0 & 13.5 & 0.69 & 0.46 \\
\hline
  2.50 -- 2.82 &  32,500 &  11.0 & 16.0 & 13.5 & 0.73 & 0.42 \\
       family  &  14,261 &  11.0 & 16.0 & 13.5 & 0.67 & 0.44 \\
   background  &  18,239 &  11.0 & 15.5 & 13.5 & 0.54 & 0.57 \\
\hline
  2.82 -- 3.60 &  24,367 &  12.0 & 15.5 & 13.5 & 0.56 & 0.40 \\
       family  &   9,547 &  12.0 & 15.0 & 13.5 & 0.57 & 0.37 \\
   background  &  14,820 &  12.0 & 15.0 & 13.5 & 0.56 & 0.52 \\
\hline
\end{tabular}
\\
\vspace{1mm}
 {
 \scriptsize
 {\it Notes:}\\
$^a$ the range of proper semi-major axis for defining the inner,
middle and outer main belt (AU);  \\
$^b$ the number of objects in each subsample; the first line corresponds
to the full sample, and the following two to subsamples classified as
families and background, respectively. The total number of objects
is 87,569. \\
$^c$ the minimum and maximum $H$ magnitude used in fitting the $H$
distribution \\
$^d$ the best-fit ``break'' $H$ magnitude (see text) \\  
$^e$ the ``bright'' $H$ distribution slope \\
$^f$ the ``faint'' $H$ distribution slope \\
}
\caption{Best-fit parameters for counts shown in Fig.~\ref{9panelSD}}
\end{table}
%%%%%%%%%%%%%%%%%%%%%%%%

\begin{table}
\scriptsize
\centering
\begin{tabular}{|r|r|r|r|r|r|}
\hline
Family$^a$  & $N^b$ &
 $H_b^c$ & $H_f^d$ & $\alpha_S^e$ & age$^f$ \\ 
\hline
\hline
Baptistina &   310 &  14.0 &  16.0 & 0.49 & --- \\
McCuskey &  1043 &  12.5 &  16.0 & 0.49 & --- \\
Erigone &   307 &  14.0 &  16.0 & 0.59 & --- \\
Euterpe &   387 &  13.5 &  16.0 & 0.52 & --- \\
Andree &   649 &  14.0 &  16.0 & 0.70 & --- \\
Massalia &   730 &  14.5 &  16.0 & 0.97 & 0.3 $\pm$ 0.1 \\
Rafita &   225 &  14.0 &  16.0 & 0.35 & 1.5 $\pm$ 0.5 \\
Mitidika &   698 &  12.5 &  15.5 & 0.61 & --- \\
Misa &   185 &  13.0 &  15.5 & 0.47 &  0.5 $\pm$ 0.2 \\
Juno &   354 &  14.5 &  16.0 & 0.63 & --- \\
Aeolea &   172 &  14.0 &  15.5 & 0.85 & --- \\
Nemesis &   129 &  14.0 &  16.0 & 0.70 & 0.2 $\pm$ 0.1 \\
Lydia &   598 &  14.0 &  16.0 & 0.81 & --- \\
Padua &   442 &  14.0 &  16.0 & 0.50 & --- \\
Merxia &   252 &  14.5 &  15.5 & 0.71 & 0.5 $\pm$ 0.2 \\
Chloris &   121 &  13.0 &  15.5 & 0.35 &  0.7 $\pm$ 0.4 \\
Agnia &  1106 &  13.0 &  15.5 & 0.52 &  0.2 $\pm$ 0.1 \\
Brasilia &   127 &  13.5 &  15.5 & 0.71 & 0.05 $\pm$ 0.04 \\
Lixiaohua &   150 &  13.0 &  15.0 & 0.56 & 0.3 $\pm$ 0.2 \\
Ursula &   644 &  11.0 &  15.0 & 0.47 & --- \\
Theobalda &   100 &  13.0 &  15.5 & 0.44 & --- \\
Veritas &   250 &  12.0 &  15.0 & 0.50 & 8.3 $\pm$ 0.5 Myr \\
\hline
\end{tabular}
\\
\vspace{1mm}
{
 \scriptsize
 {\it Notes:}\\
$^a$ the family name (the lowest-numbered member) \\
$^b$ the number of objects in SDSS MOC 4 associated with this family \\
$^c$ bright $H$ magnitude limit used for fitting \\
$^d$ faint $H$ magnitude limit used for fitting \\
$^e$ the best-fit power-law index for $H$ distribution \\
$^e$ the family age in Gyr (except for Veritas in Myr), taken from
     Nesvorn\'{y} et al. (2005), when available  \\
}
\caption{Best-fit $H$ distribution parameters for families described by
a single power law.}
\end{table}
%%%%%%%%%%%%%%%%%%%%%%%%

\begin{table}
\scriptsize
\centering
\begin{tabular}{|r|r|r|r|r|r|r|r|}
\hline
Family$^a$  & $N^b$ &
 $H_b^c$ & $H_f^d$ & $H_{B}^e$ & $\alpha_1^f$ & $\alpha_2^g$ & age$^h$ \\ 
\hline
\hline
Flora &  6164 &  11.0 &  16.0 &  14.0 & 0.66 & 0.40 & 1.1 $\pm$ 0.5 \\
Vesta &  3793 &  12.5 &  16.0 &  14.8 & 0.89 & 0.50 &  --- \\
NysaPolana &  2928 &  13.5 &  16.5 &  15.0 & 0.80 & 0.39 &  --- \\
Teutonia &  3405 &  12.5 &  15.5 &  14.0 & 0.94 & 0.59 &  --- \\
Maria &  1315 &  11.0 &  15.5 &  14.5 & 0.53 & 0.38 & 3.0 $\pm$ 1.0 \\
Eunomia &  2995 &  11.0 &  15.5 &  13.5 & 0.76 & 0.28 & 2.5 $\pm$ 0.5 \\
Adeona &   428 &  12.0 &  15.5 &  14.2 & 0.57 & 0.29 & 0.7 $\pm$ 0.5 \\
Henan &   624 &  14.0 &  16.0 &  15.1 & 0.79 & 0.46 &  --- \\
Gefion &   914 &  12.0 &  15.5 &  14.2 & 0.86 & 0.26 & 1.2 $\pm$ 0.4 \\
Dora &   248 &  13.0 &  16.0 &  14.8 & 0.37 & 0.18 & 0.5 $\pm$ 0.2 \\
Koronis &  1267 &  12.0 &  15.0 &  13.5 & 0.55 & 0.26 & 2.5 $\pm$ 1.0 \\
Eos &  4367 &  10.5 &  15.0 &  13.9 & 0.52 & 0.32 & 2.0 $\pm$ 0.5 \\
Tirela &   411 &  13.2 &  15.0 &  14.0 & 1.04 & 0.62 &  --- \\
Themis &  1073 &  11.0 &  15.0 &  13.4 & 0.46 & 0.10 & 2.5 $\pm$ 1.0 \\
Hygiea &  1076 &  12.7 &  15.0 &  14.4 & 0.62 & 0.18 & 2.0 $\pm$ 1.0 \\
\hline
\end{tabular}
\\
\vspace{1mm}
 {
 \scriptsize
 {\it Notes:}\\
$^a$ the family name (the lowest-numbered member) \\
$^b$ the number of objects in SDSS MOC 4 associated with this family \\
$^c$ bright $H$ magnitude limit used for fitting \\
$^d$ faint $H$ magnitude limit used for fitting \\
$^e$ the ``break'' $H$ magnitude used for fitting \\
$^f$ the best-fit power-law index for $H$ distribution at the bright end\\
$^g$ the best-fit power-law index for $H$ distribution at the faint end\\
$^h$ the family age in Gyr, taken from
     Nesvorn\'{y} et al. (2005), when available  \\
}
\caption{Best-fit $H$ distribution parameters for families described by
a ``broken'' power law.}
\end{table}
%%%%%%%%%%%%%%%%%%%%%%%%

}

\end{document}